\documentclass[a4paper,twocolumn,11pt, accepted=2023-12-03]{quantumarticle}
\pdfoutput=1
\usepackage[utf8]{inputenc}
\usepackage[english]{babel}
\usepackage[T1]{fontenc}
\usepackage{amsthm}
\usepackage{amssymb}
\usepackage{amsmath}
\usepackage{mathtools}
\usepackage[ruled,linesnumbered]{algorithm2e}
\usepackage{hyperref}
\usepackage{xspace}

\usepackage{physics}
\usepackage{xcolor}
\usepackage{siunitx}
\usepackage{multirow}
\usepackage{booktabs}

\usepackage[numbers]{natbib}

\theoremstyle{definition}
\newtheorem{definition}{Definition}[section]

\newcommand{\cz}{\textsc{cz}\xspace}
\newcommand{\mr}[1]{\mathrm{#1}}
\newcommand{\figineq}[2]{\vcenter{\hbox{\includegraphics[width=#1]{#2}}}}
\newcommand{\adj}{\operatorname{adj}\xspace}
\newcommand{\stargraph}[1]{G_*^{\qty(#1)}}
\newcommand{\Exp}[1]{\mathbb{E}\qty[#1]}
\newcommand{\FT}[1]{\mathcal{F}\qty[#1]}
\newcommand{\psucc}{p_\mr{succ}}

\newcommand{\revise}[1]{{#1}}

\SetKwComment{Comment}{ // }{}

\begin{document}

\title{Graph-theoretical optimization of fusion-based graph state generation}

\author{Seok-Hyung Lee}
\affiliation{Department of Physics and Astronomy, Seoul National University, Seoul 08826, Republic of Korea}
\affiliation{Centre for Engineered Quantum Systems, School of Physics, University of Sydney, Sydney, NSW 2006, Australia}
\email{seokhyung.lee@sydney.edu.au}
\orcid{0000-0002-1207-2752}
\author{Hyunseok Jeong}
\email{h.jeong37@gmail.com}
\affiliation{Department of Physics and Astronomy, Seoul National University, Seoul 08826, Republic of Korea}
\maketitle

\begin{abstract}
  Graph states are versatile resources for various quantum information processing tasks, including measurement-based quantum computing and quantum repeaters. 
  Although the type-II fusion gate enables all-optical generation of graph states by combining small graph states, its non-deterministic nature hinders the efficient generation of large graph states.
  In this work, we present a graph-theoretical strategy to effectively optimize fusion-based generation of any given graph state, along with a Python package \textit{OptGraphState}.
  Our strategy comprises three stages: simplifying the target graph state, building a fusion network, and determining the order of fusions. 
  Utilizing this proposed method, we evaluate the resource overheads of random graphs and various well-known graphs.
  \revise{Additionally, we investigate the success probability of graph state generation given a restricted number of available resource states.}
  We expect that our strategy and software will assist researchers in developing and assessing experimentally viable schemes that use photonic graph states.
\end{abstract}

\emph{Graph states} represent a family of multi-qubit states where qubits are entangled following a specific structure determined by an associated graph. 
Owing to their highly entangled nature \cite{hein2006entanglement}, graph states find applications in various quantum information processing domains, such as measurement-based quantum computing (MBQC) \cite{raussendorf2001one,raussendorf2003measurement, raussendorf2006fault, raussendorf2007topological}, fusion-based quantum computing (FBQC) \cite{bartolucci2023fusion}, quantum error correction \cite{schlingemann2001quantum,pirker2017construction}, quantum secret sharing \cite{markham2008graph,bell2014experimental}, quantum repeaters \cite{zwerger2012measurement,zwerger2013universal,azuma2015all,wallnofer2016two}, and quantum metrology \cite{shettell2020graph}. 
Photonic qubit-based graph states are particularly crucial in these applications, not only because photons are predominantly used in quantum communication but also because MBQC can circumvent the need for in-line near-deterministic multi-qubit gates that are challenging in photonic systems \cite{nielsen2004optical}.

All-optical methods for constructing photonic graph states are commonly processed by merging multiple smaller graph states into a larger one using \emph{fusion operations of types I and/or II} \cite{browne2005resource}. 
The failures of these operations are heralded, presenting a significant advantage \cite{adcock2018hard} over alternative methods such as the post-selected controlled-Z (\cz) gate \cite{hofmann2002quantum,ralph2002linear} and the post-selected fusion gate \cite{browne2005resource}. 
Among the two fusion types, we focus exclusively on type II. 
This is because, assuming photodetectors with negligible dark counts, a type-I fusion could potentially transform a photon loss into an undetectable computational error \cite{li2015resource}, whereas any photon loss occurring before or during a type-II fusion can be identified.

The non-deterministic nature of fusions is a crucial consideration when investigating quantum tasks using photonic graph states.
When employing dual-rail-encoded qubits (such as polarization qubits) and restricting the setup to linear-optical devices and photodetectors, the success probability of a type-II fusion is limited to 50\% without ancillary resources \cite{braunstein1995measurement}. 
Higher success probabilities can be achieved by utilizing ancillary photons \cite{grice2011arbitrarily,ewert2014efficient}, encoded qubits \cite{lee2015nearly,lee2019fundamental}, redundant graph structures \cite{fujii2010fault,li2010fault,li2015resource}, or continuous-variable techniques \cite{jeong2001quantum,jeong2002efficient,omkar2020resource,omkar2021resource,takeda2013deterministic,zaidi2013beating}. 
Through these methods, fault-tolerant linear-optical MBQC is theoretically possible. 
For instance, our recent research verified that high photon loss thresholds of around 8\% under a uniform photon loss model can be attained by employing parity-encoded multiphoton qubits \cite{lee2023parity}.

Despite these advancements, resource overhead remains a significant challenge for generating large-scale graph states. 
Specifically, the number of required basic resource states (such as three-photon Greenberger–Horne–Zeilinger states) or optical elements like photodetectors increases exponentially as the number of fusions grows. 
Consequently, it is essential to carefully design a procedure for generating a desired graph state from basic resource states to minimize resource overhead as much as possible. 
While several prior studies \cite{gilbert2006efficient,kieling2007minimal} have addressed this issue, they only considered specific graph families and type-I fusion. 
In our previous work \cite{lee2023parity}, we proposed a partial solution for general graphs and type-II fusion using a fine-tuning strategy; however, there is still considerable scope for improvement.

In this work, we introduce a graph-theoretical strategy to effectively identify a resource-efficient method for fusion-based generation of any given graph state, building upon and generalizing the strategies presented in Ref.~\cite{lee2023parity}. 
A single trial of our strategy comprises three main stages: (i) simplifying the graph state through the process of \emph{unraveling}, (ii) constructing a \emph{fusion network} (a graph that dictates the required fusions between basic resource states), and (iii) determining the order of fusions.
A sufficient number of trials are repeated with randomness and the one with the smallest resource overhead is selected as the outcome of the strategy.
Although our approach does not guarantee the most optimal method, we provide evidence of its power and generality, making it suitable for studying various tasks involving graph states. 
Our strategy is implemented in an open-source Python package, \emph{OptGraphState}, which is publicly available on Github: \url{https://github.com/seokhyung-lee/OptGraphState}.

This paper is structured as follows:
In Sec.~\ref{sec:preliminaries}, we review the definitions and basic properties of graph states and type-II fusion.
In Sec.~\ref{sec:strategy}, we describe our optimization strategy step by step with examples.
In Sec.~\ref{sec:applications}, we compute the resource overheads of various graphs using our strategy and numerically verify its effectiveness by comparing it with alternative strategies that lacks certain stages of the original strategy.
\revise{We additionally discuss the success probability of generating a graph state given a restricted number of available basic resource states.}
We conclude with final remarks in Sec.~\ref{sec:remarks}.

\section{Preliminaries \label{sec:preliminaries}}

\subsection{Graph states and their equivalence relation \label{subsec:graph_state}}

For a given graph $G = (V,~E)$ with a vertex set $V$ and an edge set $E$, a graph state $\ket{G}_V$ on qubits placed at the vertices is defined as
\begin{align*}
  \ket{G}_V \coloneqq \prod_{\qty{v_1, v_2} \in E} \hat{U}_\mr{CZ}\qty(v_1, v_2) \bigotimes_{v \in V}\ket{+}_v,
\end{align*}
where $\hat{U}_\mr{CZ}\qty(v_1, v_2)$ is the controlled-Z (\cz) gate between the qubits at $v_1$ and $v_2$ and $\ket{+}_v$ is the state $\ket{0} + \ket{1}$ on $v$.
(We omit normalization coefficients throughout the paper unless necessary.)
The graph state $\ket{G}_V$ has the stabilizer group generated by
\begin{align*}
  \qty{\hat{S}_v \coloneqq \hat{X}_v \prod_{u \in \adj(v)} \hat{Z}_{u} ~\middle|~ v \in V},
\end{align*}
where $\hat{X}_v$ and $\hat{Z}_v$ are respectively the Pauli-X and Z operators on $v$ and $\adj(v)$ is the set of the adjacent vertices of $v$.
Namely, $\hat{S}_v \ket{G}_V = \ket{G}_V$ for every $v \in V$.

\begin{figure}[!t]
  \centering
  \includegraphics[width=\columnwidth]{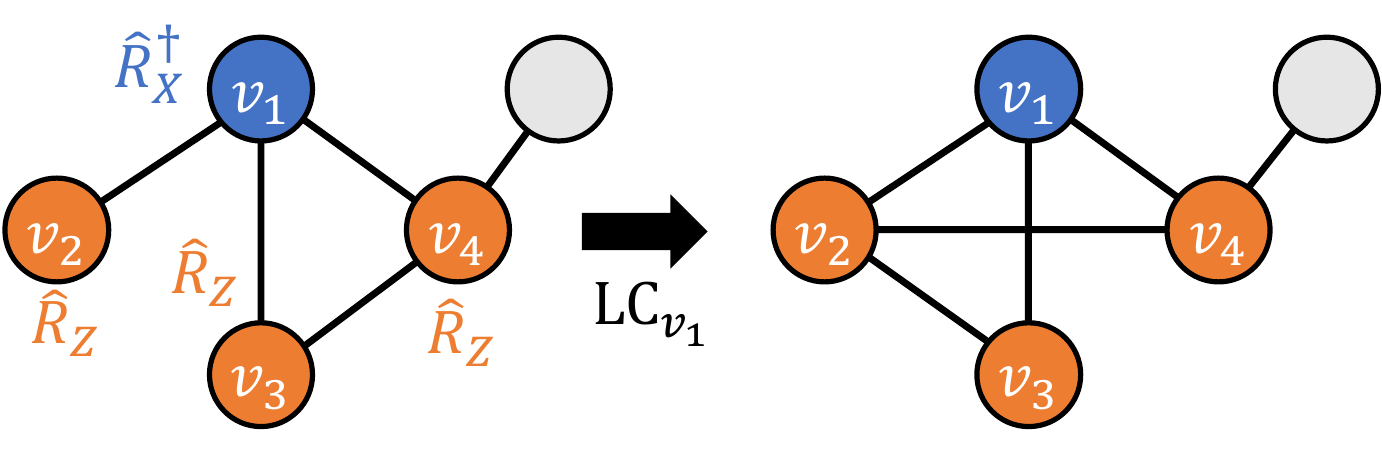}
  \caption{
    \textbf{Example of a local complementation (LC) and the corresponding single-qubit Clifford operations.}
    $\hat{R}_X$ and $\hat{R}_Z$ indicate a $\pi/2$ rotation around the $x$- and $z$-axis, respectively, in the Bloch sphere; namely, \revise{$\hat{R}_P \coloneqq \exp\qty[-i(\pi/4)\hat{P}]$} for $\hat{P} \in \qty{\hat{X}, \hat{Z}}$.
    For the presented five-qubit graph state, applying $\hat{R}_X^\dagger$ to vertex $v_1$ and $\hat{R}_Z$ to each of its neighbors is equivalent to transform the graph by an LC with respect to $v_1$.
  }
  \label{fig:local_complementation}
\end{figure}

\begin{figure*}[!t]
  \centering
  \includegraphics[width=\textwidth]{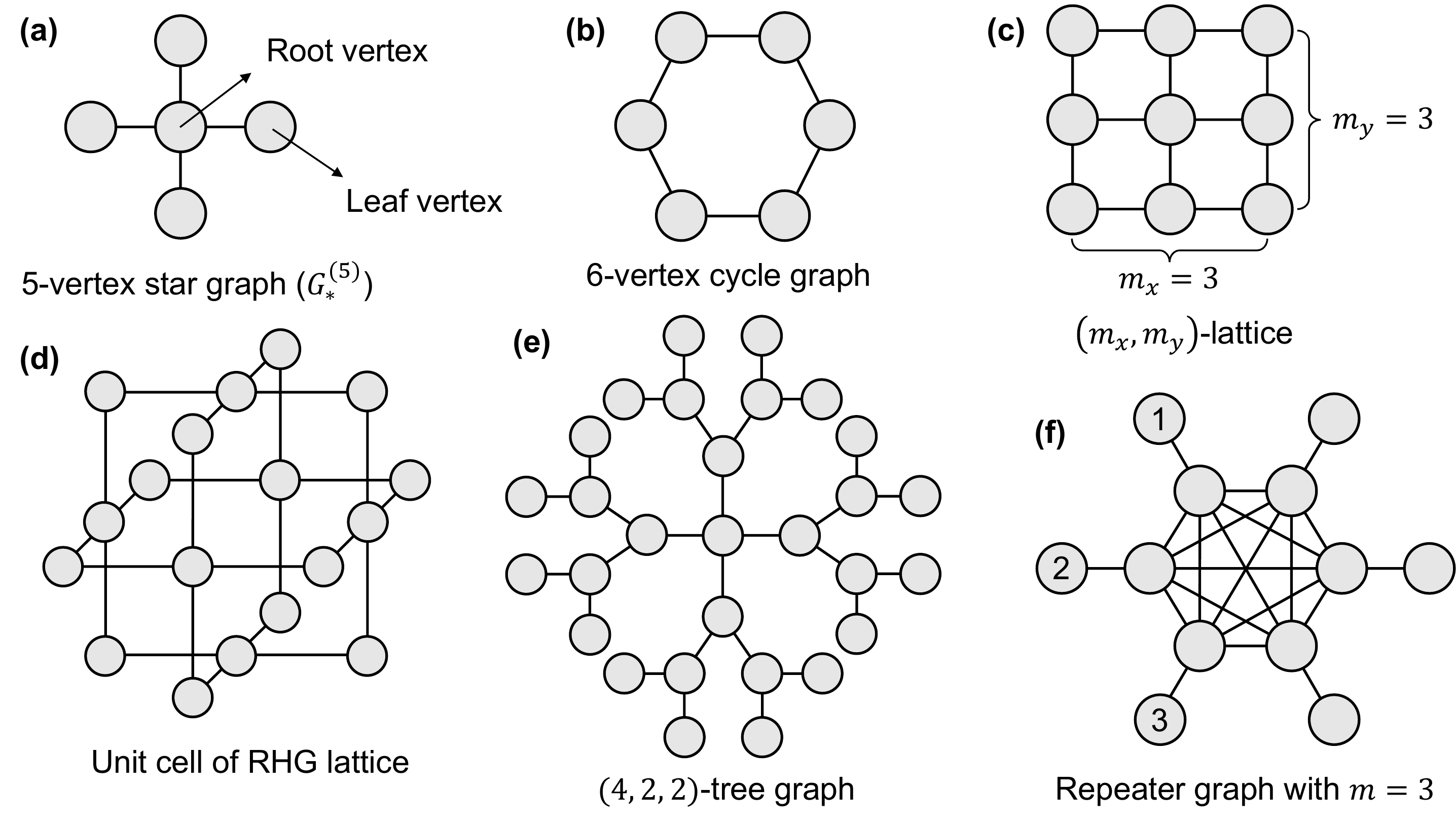}
  \caption{\textbf{Examples of various well-known families of graphs.} See Sec.~\ref{subsec:graph_state} for their descriptions and usages.}
  \label{fig:various_graphs}
\end{figure*}

An important problem regarding graph states is whether two different graph states are equivalent under a unitary operation, especially under a sequence of single-qubit Clifford operations.
For a graph $G = (V, E)$ and a vertex $v \in V$, we define a Clifford operator
\begin{align}
  \hat{U}_\mr{LC}(v) \coloneqq e^{\revise{i\frac{\pi}{4}} \hat{X}_v} \prod_{u \in \adj(v)} e^{\revise{-i\frac{\pi}{4}} \hat{Z}_u}. \label{eq:local_complementation_operator}
\end{align}
A \emph{local complementation} $\mr{LC}_v$ with respect to a vertex $v \in V$ is defined as a graph operation that, for every pair of adjacent vertices of $v$, connect them if they are disconnected and disconnect them if they are connected.
As proved in Ref.~\cite{nest2004graphical} and visualized in Fig.~\ref{fig:local_complementation}, $\hat{U}_\mr{LC}(v)$ transforms $G$ by a local complementation; namely,
\begin{align*}
  \hat{U}_\mr{LC}(v) \ket{G} = \ket{\mr{LC}_v(G)}.
\end{align*}
Furthermore, it is known that two graph states are equivalent under a sequence of single-qubit Clifford operations if and only if one of their corresponding graphs can be obtained by applying a sequence of local complementations on the other \cite{nest2004graphical}.

The followings describe several well-known families of graph states visualized in Fig.~\ref{fig:various_graphs}:

\paragraph*{Star graph.}
The $m$-vertex star graph $G^{(m)}_*$ is a graph where one of the vertices (say, $v_\mr{root}$) is connected with all the other vertices that are not connected with each other; see Fig.~\ref{fig:various_graphs}(a) for an example. 
The vertex $v_\mr{root}$ is called the \emph{root} vertex of $G^{(m)}_*$ and the other vertices are called its \emph{leaf} vertices.
Note that the graph state $\ket{G^{(m)}_*}$ is equivalent to the $m$-qubit Greenberger–Horne–Zeilinger (GHZ) state $\ket{\mr{GHZ}_m} \coloneqq \ket{0}^{\otimes m} + \ket{1}^{\otimes m}$ and the graph state of the $m$-vertex complete graph $G_\mr{cmpl}^{(m)}$ (where all the vertices are connected) under single-qubit Clifford gates; namely,
\begin{align*}
  \ket{\mr{GHZ}_m} &= \qty(\prod_{v \in V_\mr{leaf}} \hat{H}_v) \ket{G^{(m)}_*}, \\
  \ket{G_\mr{cmpl}^{(m)}} &= \hat{U}_\mr{LC}\qty(v_\mr{root}) \ket{G^{(m)}_*},
\end{align*}
where $V_\mr{leaf}$ is the set of the leaf vertices of $\stargraph{m}$ and $\hat{H}_v$ is the Hadamard gate on the qubit at $v$.
Star graph states are often used as basic resource states of photonic \revise{MBQC} \cite{li2010fault,li2015resource,omkar2020resource,omkar2022all,lee2023parity} and \revise{FBQC} \cite{bartolucci2023fusion}.

\paragraph*{Cycle graph.}
A cycle graph consists of vertices connected in a closed chain.
In particular, the graph state for the six-vertex cycle graph, which is shown in Fig.~\ref{fig:various_graphs}(b), is used as a basic resource state of FBQC \cite{bartolucci2023fusion}.

\paragraph*{Lattice graph.}
The $(m_x, m_y)$-lattice graph for integers $m_x, m_y \geq 1$ has a two-dimensional (2D) square lattice structure where the vertices are repeated $m_x$ ($m_y$) times along the $x$-axis ($y$-axis). 
See Fig.~\ref{fig:various_graphs}(b) for an example. 
Lattice graph states are particularly useful for 2D \revise{MBQC} \cite{raussendorf2001one,raussendorf2003measurement}, which is universal but not fault-tolerant \cite{raussendorf2006fault}.
Any single-qubit rotation and the controlled-\textsc{not} gate can be implemented by measuring qubits of a lattice graph state in appropriate single-qubit bases.

\paragraph*{Raussendorf-Harrington-Goyal (RHG) lattice.}
The $(L_x, L_y, L_z)$-RHG lattice graph is composed of unit cells stacked $L_x$, $L_y$, and $L_z$ times along the $x$-, $y$-, and $z$-axis, respectively.
Each unit cell is cube-shaped and it consists of vertices locating at the faces and edges of the cube, as visualized in Fig.~\ref{fig:various_graphs}(d).
RHG lattices are utilized in fault-tolerant three-dimensional (3D) MBQC \cite{raussendorf2006fault,raussendorf2007topological}.
Logical qubits encoded in a surface code can be embedded into a lattice and logical operations and measurements can be done only by single-qubit measurements and state injection.
A specific operator on each unit cell serves as a parity-check operator, whose measurement outcome is used to detect and locate errors.

\paragraph*{Tree graph.}
A tree graph is defined as a connected acyclic graph.
We particularly define the $\qty(b_0, b_1, b_2, \cdots)$-tree graph for positive integers $b_0, b_1, b_2, \cdots$ as a tree graph where a vertex (designated its \emph{root} vertex) has $b_0$ neighbors called 1st-generation \emph{branches} and each $i$th-generation branch ($i \geq 1$) has $b_i+1$ neighbors that are $(i+1)$-generation branches except for one.
As an example, see Fig.~\ref{fig:various_graphs}(e) for the $(4, 2, 2)$-tree graph.
One important usage of tree graphs is counterfactual error correction; by attaching tree graph states on qubits for 2D MBQC, qubit loss up to 50\% can be tolerated \cite{varnava2006loss}.
Such a technique also can be employed for 3D MBQC to suppress the effects of failed entangling operations during the construction of an RHG lattice \cite{li2015resource}.

\paragraph*{Repeater graph.}
The $4m$-vertex repeater graph ($m \geq 1$) consists of $2m$ completely-connected vertices and other $2m$ vertices that are respectively connected with them; see Fig.~\ref{fig:various_graphs}(f) for the case of $m=3$.
A repeater graph state can be used for all-optical quantum repeaters \cite{azuma2015all}, which distribute entanglement over a long distance by recursive entanglement swapping.
$m$ determines the number of Bell-state measurements (BSMs) required per single trial of entanglement swapping, which succeeds if any one of these $m$ BSMs succeed.

\subsection{Type-II fusion operation \label{subsec:fusion}}

\begin{figure}[!t]
  \centering
  \includegraphics[width=\columnwidth]{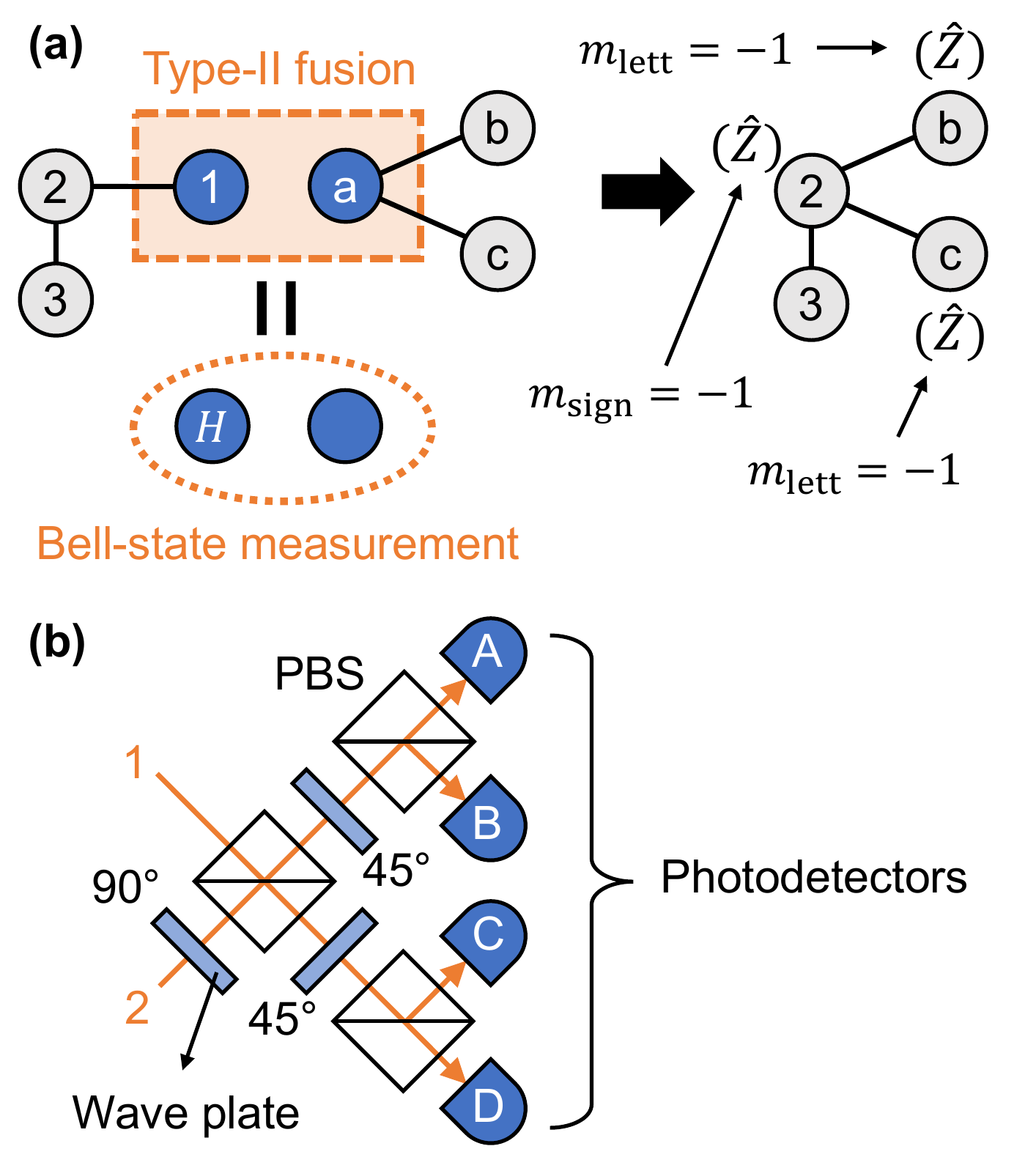}
  \caption{
    \textbf{Type-II fusion operation.}
    \textbf{(a)} Example of merging two graph states by a type-II fusion, which is composed of a Hadamard gate ($\hat{H}$) and a Bell-state measurement (BSM).
    $m_\mr{sign}$ and $m_\mr{lett}$ respectively denote the sign and letter outcomes of the BSM.
    If $m_\mr{sign}$ or $m_\mr{lett}$ is $-1$, several Pauli-Z gates need to be applied on the resulting state to get the graph state.
    \textbf{(b)} BSM scheme for single-photon polarization qubits with polarizing beam splitters (PBSs), \ang{90} and \ang{45} wave plates, and four (A--D) photodetectors.
    A PBS transmits (reflects) horizontally-polarized (vertically-polarized) photons.
    The scheme distinguishes $\ket{\psi_\pm}$: $\ket{\psi_+}$ if both A and C or both B and D detect a single photon respectively, and $\ket{\psi_-}$ if both A and D or both B and C detect a single photon respectively.
    If otherwise, the scheme fails.
    Two distinguishable Bell states can be selected by putting or removing wave plates before the first PBS.
  }
  \label{fig:fusion}
\end{figure}

The \emph{type-II fusion operation} (hereafter referred to simply as ``fusion'') \cite{browne2005resource} is a two-qubit operation that consists of applying a Hadamard gate ($\hat{H}$) to one of the qubits, followed by a BSM, and finally erasing the qubits.
In other words, fusion indicates a destructive measurement of two Pauli operators $\hat{X}\otimes \hat{Z}$ and $\hat{Z} \otimes \hat{X}$ on a pair of qubits.
By applying a fusion on an unconnected pair $\qty(v_1, v_2)$ of vertices in a graph state, we can connect (disconnect) every adjacent vertex of $v_1$ with every adjacent vertex of $v_2$ up to several Pauli-Z operators if they are unconnected (connected); see the example in Fig.~\ref{fig:fusion}(a).

More formally, for two unconnected vertices $v_1$ and $v_2$ of a graph $G$, $F_{v_1, v_2}$ is defined as a graph operation that, for every $u_1 \in \adj\qty(v_1)$ and $u_2 \in \adj\qty(v_2)$, connect (disconnect) $u_1$ and $u_2$ if they are unconnected (connected) and delete $v_1$ and $v_2$ from the graph.
When $v_1$ and $v_2$ undergo a fusion for which the Hadamard gate is applied on $v_1$, the resulting state is
\begin{align*}
  \prod_{u_1 \in \adj\qty(v_1)} \hat{Z}_{u_1}^{\frac{1 - m_\mr{sign}}{2}} \prod_{u_2 \in \adj\qty(v_2)} \hat{Z}_{u_2}^{\frac{1 - m_\mr{lett}}{2}} \ket{F_{v_1, v_2}(G)},
\end{align*}
where $(m_\mr{sign}, m_\mr{lett})$ is the outcome of the BSM.
Here, we denote the Bell basis as 
\begin{align}
  \ket{\phi_\pm} &\coloneqq \ket{00} \pm \ket{11}, \\
  \ket{\psi_\pm} &\coloneqq \ket{01} \pm \ket{10}, \label{eq:bell_basis}
\end{align}
and the outcome of a BSM as $(\pm 1, 1)$ if $\ket{\phi_\pm}$ is obtained and $(\pm 1, -1)$ if $\ket{\psi_\pm}$ is obtained.
\revise{Note that fusion was originally defined as a BSM in Ref.~\cite{browne2005resource}.
Nevertheless, we consider its variant (differing only by a Hadamard gate) since it is more suitable to generate arbitrary graph states due to its aforementioned property.}

For single-photon polarization qubits with the basis of horizontally and vertically polarized single-photon states ($\ket{\textsc{h}}$, $\ket{\textsc{v}}$), the BSM can be done with linear optical devices and photodetectors \cite{lutkenhaus1999bell}, as visualized in Fig.~\ref{fig:fusion}(b).
This BSM scheme can distinguish only two among the four Bell states, thus the fusion succeeds with the probability of
\begin{align*}
  p_{\mr{succ}}(\eta) = \frac{(1 - \eta)^2}{2}
\end{align*}
when each photon suffers loss with probability $\eta$ and the input state is maximally mixed.
See Ref.~\cite{lee2023parity} for a discussion on how failed fusions affect the resulting graph state.

\section{Strategy for identifying a method for graph state generation \label{sec:strategy}}

In this section, we present our main result: a graph-theoretical strategy to effectively identify a resource-efficient method for generating an arbtirary graph state via fusions.
Our basic resource state is the three-qubit star graph state
\begin{align}
  \ket{G^{(3)}_*} = \ket{\figineq{32pt}{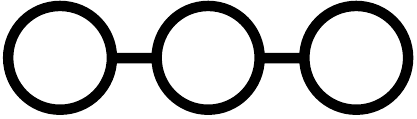}} \coloneqq \ket{+0+} + \ket{-1-},
  \label{eq:g3_state}
\end{align}
where $\ket{\pm} \coloneqq \ket{0} \pm \ket{1}$.
Hence, our goal is to find an efficient way to build a desired graph state $\ket{G}$ by performing fusions on multiple $\ket{G^{(3)}_*}$ states.
\revise{We take an approach that, whenever a fusion fails, we discard the photons that were entangled with the photons involved in the failed fusion, regenerate the state of the discarded photons with new photons, and retry the fusion.
This process does not necessarily have to proceed in order; that is, we may parallelly generate multiple identical states and post-select only the ones with successful fusions.
}

The resource overhead of a method to generate $\ket{G}$ is quantified by the \revise{expected value $Q$ of the} number of $\ket{G^{(3)}_*}$'s required to generate one $\ket{G}$ state.
\revise{For example, a $\ket{G^{(4)}_*}$ state can be generated by fusing two $\ket{G^{(3)}_*}$'s, thus this process has the resource overhead of four if $\psucc = 1/2$.
Fusing it again with another $\ket{G^{(3)}_*}$ state gives the resource overhead of $2\times(4+1)=10$.
In reality, the success probability may vary for each fusion due to different lengths of delay lines, but we simplify the problems by neglecting this fact in the following investigations.
Note that, by definition, having $Q$ basic resource states does not guarantee the successful generation of the graph state.
The exact success probability is worth investigating, which will be covered later in Sec.~\ref{subsec:generation_success_prob}.
}

\revise{The basic resource state $\ket{\stargraph{3}}$} can be generated with a success rate of \revise{$1/32$} (in a lossless case) by using linear optical devices, single-photon sources, and photodetectors \cite{varnava2008how}.
Furthermore, its deterministic generation is possible with matter-based methods \cite{schon2005sequential,lindner2009proposal}, which is experimentally demonstrated in several recent works \cite{schwartz2016deterministic,takeda2019ondemand,thomas2022efficient}.

The key concept of our strategy is a \emph{fusion network}, which is a graph where vertices correspond to individual $\ket{G^{(3)}_*}$ states and edges indicate fusions between the states required to generate $\ket{G}$.
Since fusion networks are not unique, selecting an appropriate fusion network is the first challenge.
The second challenge is to determine the order of the fusions; although the final state is regardless of the order as long as all the fusions succeed, the non-deterministic nature of fusions makes it severely affect the resource efficiency.

The strategy is summarized as follows:
\begin{enumerate}
  \item Simplify the graph of the desired graph state by \emph{unraveling} subgraphs of specific types (bipartitely-complete subgraphs and cliques). (Sec.~\ref{subsec:unraveling})
  \item Construct a fusion network from the simplified graph by decomposing it into multiple star graphs and replacing each star graph with multiple $\ket{G^{(3)}_*}$ states. (Sec.~\ref{subsec:fusion_network})
  \item Determine the fusion order with the \emph{min-weight-maximum-matching-first} method. (Sec.~\ref{subsec:fusion_order})
  \item Iterate the above steps (which contain randomness) a sufficient number of times and select the best one. (Sec.~\ref{subsec:iteration})
\end{enumerate}
We cover these steps in the following four subsections one by one.

\subsection{Simplification of graph by unraveling \label{subsec:unraveling}}

If the graph $G = (V, E)$ of the desired graph state $\ket{G}$ contains specific types of subgraphs, it is posible to generate $\ket{G}$ by applying single-qubit Clifford operations and/or fusions on the graph state of a simplified graph.
\emph{Unraveling} means the process to build such a simplified graph $G_\mr{unrv}=\qty(V', E')$ (referred to as an \emph{unraveled graph}) and specify the information $\qty(\hat{U}_\mr{C}, \mathcal{F})$ necessary to recover $\ket{G}$ from $\ket{G_\mr{unrv}}$, where $\hat{U}_\mr{C}$ is the product of single-qubit Clifford operations and $\mathcal{F} \subset V' \times V'$ is the set of pairs of vertices that undergo fusions.
We currently have unraveling schemes for two types of subgraphs: bipartitely-complete subgraphs and cliques.

\subsubsection{Unraveling bipartitely-complete subgraphs \label{subsubsec:unraveling_cbs}}

\begin{figure}[!t]
  \centering
  \includegraphics[width=\columnwidth]{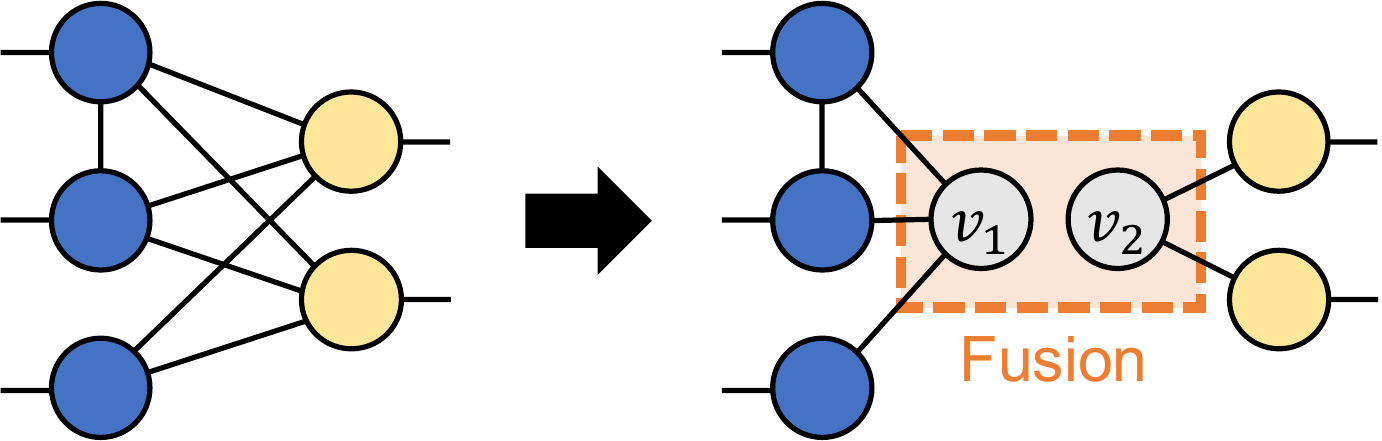}
  \caption{
    \textbf{Example of an unraveling process of a bipartitely-complete \revise{subgraph}.}
    The vertices of the two parts are colored in blue and yellow, respectively.
    The original graph state can be constructed by performing a fusion on $v_1$ and $v_2$ of the unraveled graph state.
  }
  \label{fig:unraveling_bcs}
\end{figure}

\begin{definition}[\textbf{Bipartitely-complete graph/subgraph}]
  A graph $G = (V, E)$ is an $(n, m)$ \emph{bipartitely-complete graph} for an integer $n, m \geq 2$ if $V$ can be split into two disjoint subsets $V_1$ and $V_2$ such that $\abs{V_1}=n$, $\abs{V_2}=m$, and each vertex of $V_1$ is connected with each vertex of $V_2$ (namely, $\qty{v_1, v_2} \in E$ for any $v_1 \in V_1$ and $v_2 \in V_2$).
  If a subgraph of a graph is an $(n, m)$ bipartitely-complete graph, it is called an $(n, m)$ \emph{bipartitely-complete subgraph (BCS)}.
  \label{def:bipartitely_complete_graph}
\end{definition}

Note that a bipartitely-complete graph allows edges between vertices in one part, thus it is different from a complete bipartite graph.
The parameter $(n, m)$ of a bipartitely-complete graph may not be uniquely determined.

If $G$ has an $(n, m)$ BCS, its two parts can be disconnected by adding two vertices $\qty(v_1, v_2)$ that are respectively connected with all the vertices in one of the two parts and adding $\qty(v_1, v_2)$ to $\mathcal{F}$, which replace $nm$ edges with $n+m$ edges and one fusion; see Fig.~\ref{fig:unraveling_bcs} for an example.
This process is called unraveling the BCS.

In our strategy, we repeat the cycle of finding non-overlapping BCSs (that do not share any vertices) via Algorithm~\ref{alg:finding_bcss} and unraveling them as above until no new BCSs are found.
The time complexity of Algorithm~\ref{alg:finding_bcss} is $\order{|V|d_\mr{max}^4}$ in the worst case, where $d_\mr{max}$ is the largest degree\footnote{The degree of a vertex is the number of its adjacent vertices.}.
Note that the iterations in Algorithm~\ref{alg:finding_bcss} are done in random orders because the final unraveled graph may vary depending on the orders and we want to suppress any possible bias during iteration (Step~4 of the strategy).
All the randomness appearing from now on exist for the same reason.

\begin{algorithm*}[t]
\caption{Finding non-overlapping bipartitely-complete subgraphs (BCSs)}
\label{alg:finding_bcss}
\KwIn{A graph $G = (V, E)$.}
\KwOut{A set $\mathcal{S}$ of BCSs of $G$ that do not share any vertices.}
$\mathcal{S} \gets \emptyset$\;
$V_\mr{in.bcs} \gets \emptyset$\Comment*[l]{Vertices included in any BCS found so far}
$E_\mr{checked} \gets \emptyset$\Comment*[l]{Edges checked so far}
\ForEach{$v \in V$ (in a random order)}{
  \uIf{$v \notin V_\mr{in.bcs}$}{
    $V_\mr{adj.unchecked} \gets \qty{v_\mr{adj} \in \adj(v) ~\middle|~ \qty{v, v_\mr{adj}} \notin E_\mr{checked}}$\;
    \ForEach{unordered pair $\qty{v_1, v_2}$ of elements in $V_\mr{adj.unchecked}$ (in a random order)}{
      \uIf{$v_1,v_2 \notin V_\mr{in.bcs}$}{
        $V_1 \gets \adj\qty(v_1) \cap \adj\qty(v_2)$\Comment*[l]{First part of BCS, which includes $v$}
        $V_2 \gets \bigcap_{u \in V_1} \adj(u)$\Comment*[l]{Second part of BCS, which includes $v_1$ and $v_2$}
        \uIf{$\abs{V_1} > 1$ and $\abs{\qty(V_1 \cup V_2) \cap V_\mr{in.bcs}} = 0$}{
          $\mathcal{S} \gets \mathcal{S} \cup \qty{\qty(V_1, V_2)}$\;
          $V_\mr{in.bcs} \gets V_\mr{in.bcs} \cup V_1 \cup V_2$\;
        }
      }
    }
    \ForEach{$v_\mr{adj} \in \adj(v) \setminus V_\mr{in.bcs}$}{
      $E_\mr{checked} \gets E_\mr{checked} \cup \qty{\qty{v, v_\mr{adj}}}$\;
    }
  }
}
\end{algorithm*}

\subsubsection{Unraveling cliques \label{subsubsec:unraveling_cliques}}

\begin{figure*}[!t]
  \centering
  \includegraphics[width=\textwidth]{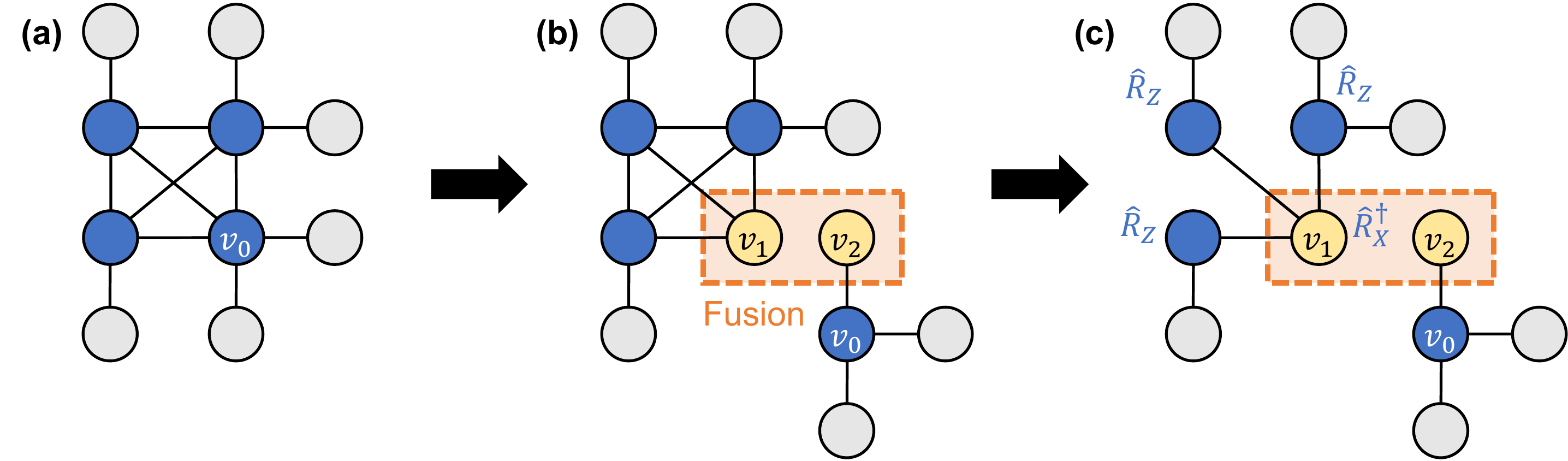}
  \caption{
    \textbf{Example of an unraveling process of a maximal clique.}
    The process is done through three steps: \textbf{(a)} selecting a random vertex $v_0$ in the clique, \textbf{(b)} \revise{separating $v_0$ from the clique} by adding two vertices $v_1$ and $v_2$, and \textbf{(c)} applying a local complementation with respect to \revise{$v_1$}.
    Vertices in the clique are colored in blue and the new vertices are colored in yellow.
    The original graph state can be constructed by performing single-qubit Clifford operations (\revise{$\hat{R}_P \coloneqq \exp\qty[-i(\pi/4)\hat{P}]$} for $\hat{P} \in \qty{\hat{X}, \hat{Z}}$) and a fusion on the unraveled graph state in \textbf{(c)}.
  }
  \label{fig:unraveling_clique}
\end{figure*}

\begin{definition}[\textbf{Clique}]
  A \emph{clique} of a graph $G$ is a subgraph of $G$ where every vertex is fully connected with each other.
  A clique is \emph{maximal} if it cannot be enlarged by adding a new vertex.
\end{definition}

If $G$ contains a clique, it can be simplified by using a local complementation.
For a maximal clique of size greater than two, the unraveling process is conducted as follows (see Fig.~\ref{fig:unraveling_clique} for an example):
\begin{enumerate}
  \item Let us define $V_\mr{cl}$ as the set of the vertices in the clique and $V_\mr{no.outer} \subseteq V_\mr{cl}$ as the set of the vertices in the clique that are connected only with vertices in the clique.
  \item If $V_\mr{no.outer}$ is not empty, select a vertex $v_0$ randomly from $V_\mr{no.outer}$.
  \item If $V_\mr{no.outer}$ is empty, \revise{do:}
  \revise{
  \begin{enumerate}
    \item Select a vertex $v_0$ randomly from $V_\mr{cl}$.
    \item Separate $v_0$ and its neighbors that are not in $V_\mr{cl}$ from the clique and add a new vertex $v_1$ at the original position of $v_0$. 
    \item Add another new vertex $v_2$ and connect it with $v_0$.
    \item Add $\qty(v_1, v_2)$ to $\mathcal{F}$.
  \end{enumerate}
  }
  \item Transform the graph by a local complementation with respect to $v_1$ and update $\hat{U}_\mr{C} \gets \hat{U}_\mr{LC}\qty(v_1) \hat{U}_\mr{C}$.
\end{enumerate}

In our strategy, we repeat the cycle of finding non-overlapping maximal cliques (that do not share any vertices) and unraveling them as above until no new cliques are found.
Listing all maximal cliques of a graph is an important problem in the graph theory and known to take exponential time in the worst case \cite{moon1965cliques}.
However, there exist algorithms to list them in polynomial time per clique \cite{lawler1980generating,tsukiyama1977new}, thus the problem can be efficiently solved if the graph does not contain many cliques.
Our Python package \href{https://github.com/seokhyung-lee/OptGraphState}{\textit{OptGraphState}} uses {\tt Graph.maximal\_clique} method from Python package \textit{python-igraph} \cite{csardi2006igraph}, which implements a more advanced algorithm in Ref.~\cite{eppstein2010listing}.

\subsubsection{Additional notes}

It is a subtle problem which of bipartitely-complete graphs and cliques to unravel first.
We randomly choose it since we currently have no basis for judging which one is better.

\begin{figure*}[!t]
  \centering
  \includegraphics[width=\textwidth]{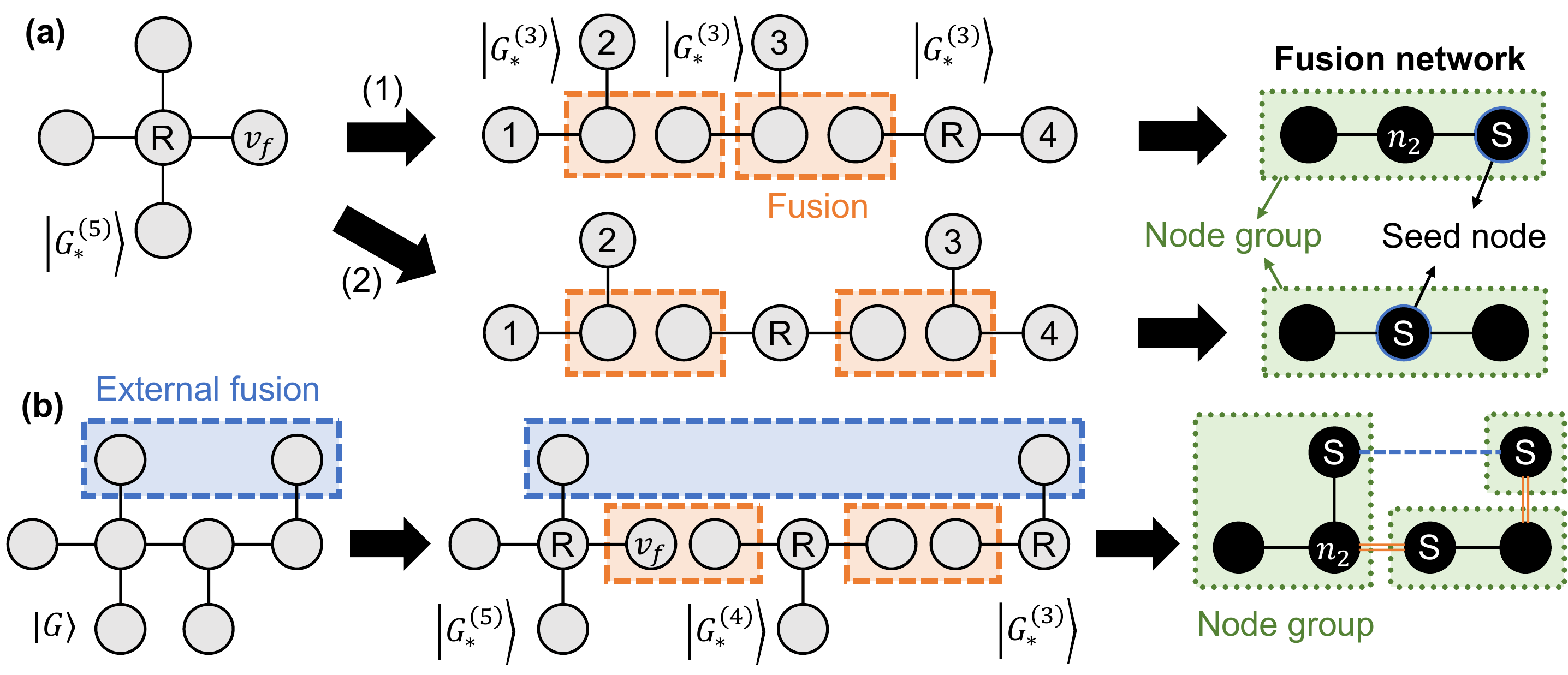}
  \caption{
    \textbf{Examples of the construction of fusion networks.}
    \textbf{(a)} A five-qubit star graph state $\ket{G^{(5)}_*}$ and \textbf{(b)} a general graph state $\ket{G}$ are considered.
    In \textbf{(a)}, $\ket{G^{(5)}_*}$ is decomposed into three $\ket{G^{(3)}_*}$ states, which leads to a three-node linear fusion network that forms one node group.
    The process varies depending on the selection of the seed node (marked as ``S''), which determines the root vertex of $G^{(5)}_*$ (marked as ``R'').
    A leaf vertex $v_f$ of the star graph can be any of the four vertices (1--4) after the decomposition.
    In \textbf{(b)}, $\ket{G}$ is decomposed into multiple star graph states, where each of them is again decomposed into $\ket{G^{(3)}_*}$ states and forms one node group in the fusion network.
    The line styles of the links in the fusion network indicate their origins: black solid lines for fusions inside a star graph, orange double lines for fusions between star graphs, and a blue dashed line for an external fusion.
  }
  \label{fig:fusion_network}
\end{figure*}
One may expect that BCSs and cliques are quite non-trivial and not very common.
However, the smallest BCS and clique that are concerned for unraveling are cycles with four vertices and with three vertices, respectively.
These are simple enough and appear in various graphs such as square and triangular grid graphs and RHG lattices.
Moreover, large bipartitely-complete subgraphs may appear when converting a logically encoded graph state into a graph state with physical qubits.
For instance, a three-qubit linear graph state with the $(n, m)$ parity encoding contains at most $(n, m)$ bipartitely-complete subgraphs in the physical level; see Sec.~\ref{subsec:well_known_graphs} and Ref.~\cite{lee2023parity} for more details.

\subsection{Construction of fusion network \label{subsec:fusion_network}}

After unraveling a graph $G$, we obtain an unraveled graph $G_\mathrm{unrv}$ along with the information $\qty(\hat{U}_\mr{C}, \mathcal{F})$ that identifies the operations needed to to restore $\ket{G}$ from $\ket{G_\mr{unrv}}$.
In particular, the fusions specified by $\mathcal{F}$ are called \emph{external fusions} to distinguish them from the fusions used to generate the unraveled graph state.
We now deal with the problem of building a fusion network from result.

We first formally define fusion networks as follows:

\begin{definition}[\textbf{Fusion network}]
  A graph $\mathcal{N}_f = (N, L)$ is a \emph{fusion network} of a graph state $\ket{G}$ (where vertices and edges are refereed to as \emph{nodes} and \emph{links}) for \emph{root indicators} $\qty{r_{l,n} \in \qty{0, 1} ~\middle|~ \forall l \in L,~ \forall n \in l}$ if $\ket{G}$ can be generated by the process:
    \begin{enumerate}
      \item Prepare a state $\ket{G^{(3)}_*}$ for each node $n$.
      Let $q_\mr{root}^{(n)}$ denote its root qubit and $Q_\mr{leaf}^{(n)}$ denote the set of its leaf qubits.
      \item For each link $l=\qty{n_1, n_2}$, iterate the following:
      \begin{enumerate}
        \item Let $q_1$ be $q_\mr{root}^{\qty(n_1)}$ if $r_{l,n_1}=1$ and an arbitrary unmeasured qubit in $Q_\mr{leaf}^{(n_1)}$ if $r_{l,n_1}=0$. Define $q_2$ analogously for $n_2$.
        \item Apply appropriate single-qubit Clifford operations on $q_1$ and $q_2$, if required.
        \item Perform a fusion on $q_1$ and $q_2$.
      \end{enumerate}
      \item Apply appropriate single-qubit Clifford operators on the remaining qubits, if required.
    \end{enumerate}
  We say that a link $l = \qty{n_1, n_2}$ has the type of \emph{root-to-root}, \emph{root-to-leaf}, or \emph{leaf-to-leaf}, when both $r_{l, n_1}$ and $r_{l, n_2}$ are equal to 1, only one of them is equal to 1, and both of them are equal to 0, respectively.
  \label{def:fusion_network}
\end{definition}

We now describe how to build a fusion network $\mathcal{N}_f$ and the corresponding root indicators $\qty{r_{l,n}}$ from the unraveled graph $G_\mr{unrv}$ and the external fusions $\mathcal{F}$.
The main idea is to decompose a graph state into multiple star graph state, each of which is again decomposed into multiple $\ket{G^{(3)}_*}$ states.

An $m$-qubit star graph state $\ket{G^{(m)}_*}$ can be constructed by conducting fusions on $m - 2$ copies of $\ket{G^{(3)}_*}$, which leads to a fusion network with $m-2$ nodes connected linearly with root-to-leaf links; see Fig.~\ref{fig:fusion_network}(a) for an example when $m=5$.
Note that there is an ambiguity in positioning the root qubit (marked as ``R'') of $\ket{G^{(m)}_*}$ as depicted in Fig.~\ref{fig:fusion_network}(a) with (1) and (2).
The node for the $\ket{G^{(3)}_*}$ state containing the root qubit of $\ket{G^{(m)}_*}$ is called the \emph{seed node} (marked as ``S'') of the \textit{node group} that consists of these $m-2$ nodes.

A general graph state $\ket{G}$ can be generated by conducting fusions on leaf qubits of multiple star graph states, where each star graph is originated from a vertex in $G$ with degree larger than one.
Consequently, its fusion network can be constructed by connecting the fusion networks of the individual star graphs (which respectively form one node group) with leaf-to-leaf links.
An example is illustrated in Fig.~\ref{fig:fusion_network}(b), where root-to-leaf links and leaf-to-leaf links are represented by black single lines and orange double lines, respectively.
If an external fusion exists, it also creates a link (blue dashed line) between nodes of different star graphs, as shown in Fig.~\ref{fig:fusion_network}(b).
Such a link may belong to any one of the three types, depending on the vertices involved in the external fusion.
Note that external fusions always appear between different star graphs, considering the unraveling processes in Figs.~\ref{fig:unraveling_bcs} and \ref{fig:unraveling_clique}.

It is important that the above process contains two types of ambiguity for each star graph (which are determined randomly in our strategy): which node in the node group to select as its seed node and which node to include each leaf vertex in.
To illustrate the latter factor with the example of Fig.~\ref{fig:fusion_network}(a), the leaf vertex $v_f$ in $\stargraph{5}$ can be any of the four vertices (1--4) after the decomposition.
Such ambiguity matters if $\stargraph{5}$ appears during the decomposition of a larger graph state.
In other words, if $v_f$ participates in a fusion, the resulting fusion network may vary depending on this selection.
For example, $\stargraph{5}$ appears in the decomposition of Fig.~\ref{fig:fusion_network}(b), and in this case, vertex~3 in (1) is selected to be $v_f$; thus, the link for the fusion is connected to node $n_2$.

We lastly note that the single-qubit Clifford operators required in the process of generating a graph state can be identified from $\hat{U}_\mr{C}$, the product of single-qubit Clifford operations obtained from the unraveling process, and the fusion outcomes.

\subsection{Determination of fusion order \label{subsec:fusion_order}}

We now have one stage left: how to determine the order of fusions.
Let us regard a fusion network as a weighted graph where each node indicates a group of entangled qubits and each link represents a fusion between these groups that needs to be done.
The weight of each node $w(n)$, which is initialized to 1, is defined as the resource overhead of the process of generating the corresponding entangled states.
Namely, $w(n)$ is the average number of required $\ket{G^{(3)}_*}$ states to generate the state.

Upon the above setting, the action of a fusion can be treated as the \emph{contraction} of a link $l$, which means to merge two endpoints $\qty(n_1, n_2)$ of the link into a new node $n_l$ and reconnect the links connected to the original nodes with $n_l$.
The weight of $n_l$ is updated as
\begin{align}
  w\qty(n_l) \coloneqq \begin{cases}
    \qty[w\qty(n_1) + w\qty(n_2)]/{\psucc} & \text{if } n_1 \neq n_2, \\
    w\qty(n_1)/{\psucc} & \text{if } n_1 = n_2,
  \end{cases}
  \label{eq:contraction_weight_update}
\end{align}
where $\psucc$ is the fusion success probability.
Hence, if the order of the fusions is given, the resource overhead $Q$ of the entire process can be obtained as the summation of the weights of the last remaining nodes after contracting all the edges in the order.
Note that an intermediate fusion network during this process may have loops (links connecting a node to itself) or multi-links (links incident to the same two vertices).

For a fusion network $\mathcal{N}_\mr{f} = (N, L)$, the number of possible fusion orders is $\abs{L}!$; thus it is extremely inefficient to randomly sample fusion orders and find the best one unless there are very few links.
Instead of it, our strategy is based on the following two intuitions:
\begin{enumerate}
  \item It is preferred to contract links with small weights first, where the weight of a link $l$ is defined as $w\qty(n_l)$ in Eq.~\eqref{eq:contraction_weight_update}. 
  It is because, defining $f(x,y) \coloneqq \qty(x + y)/{\psucc}$ for two numbers $x$ and $y$, 
    \begin{align*}
      f\qty(f\qty(w_1,w_2),w_3) < f\qty(w_1,f\qty(w_2,w_3))
    \end{align*}
  when $w_1 < w_2 < w_3$.
  \item Links that do not share endpoints can be contracted simultaneously and it is preferred to contract links as parallelly as possible.
  For example, let us consider a four-node linear fusion network where the node set is $\qty{n_1, n_2, n_3, n_4}$ and $n_i$ and $n_{i+1}$ are connected with a link $l_{i,i+1}$ for each $i \in \qty{1, 2, 3}$.
  Provided that $\psucc=1/2$, we obtain $Q=16$ if $l_{1,2}$ and $l_{3,4}$ are first contracted in parallel, but obtain $Q=22$ if $l_{2,3}$ is contracted first.
\end{enumerate}

\begin{figure}[!t]
  \centering
  \includegraphics[width=\columnwidth]{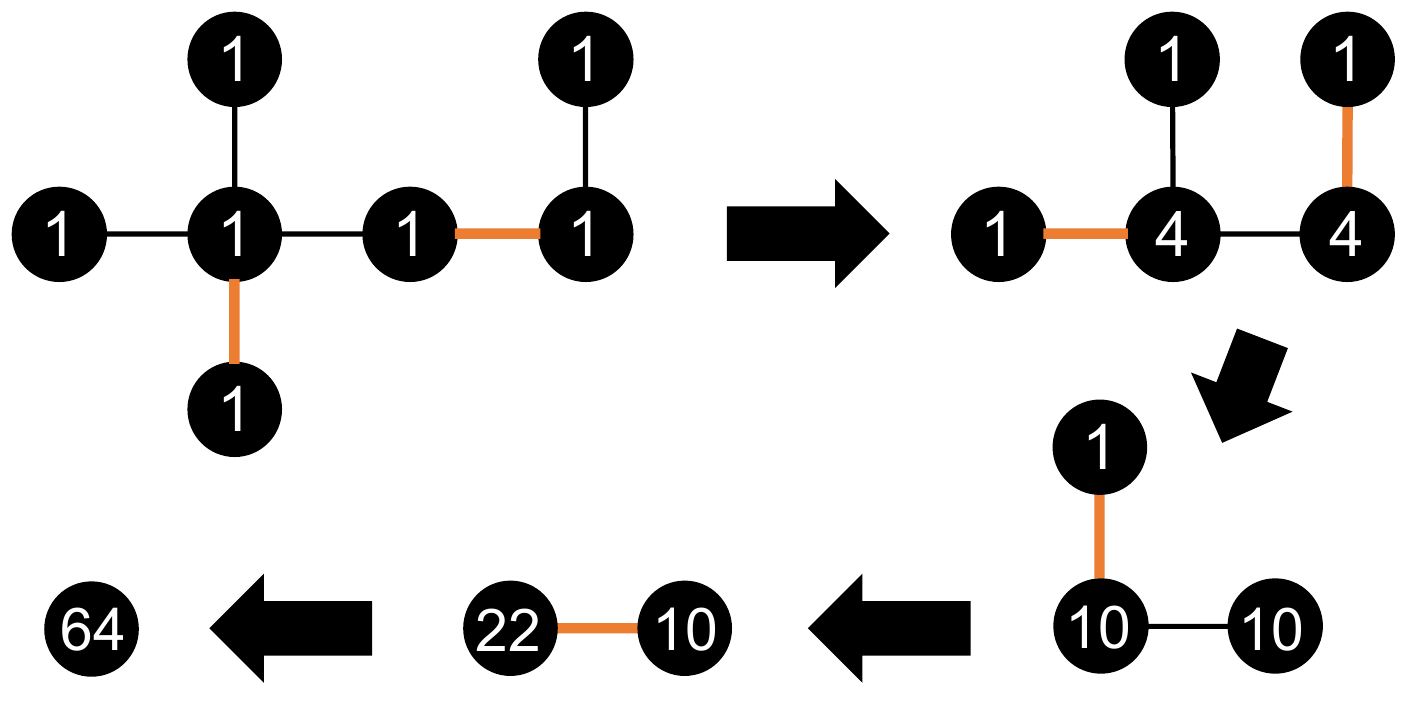}
  \caption{
    \textbf{Example of the determination of the fusion order with the min-weight-maximum-matching-first method.}
    We assume $\psucc=1/2$. 
    Each step is an intermediate fusion network after contracting links (orange bold lines) in the previous step.
    The numbers inside the nodes indicate their weights.
    The obtained resource overhead is $Q=64$, which is the weight of the last remaining node.
  }
  \label{fig:fusion_order}
\end{figure}

Based on these intuitions, we introduce the \emph{min-weight-maximum-matching-first} method to determine the fusion order.
For each round of the link contraction process, we first identify the set of links with the smallest weight and get the subgraph $\mathcal{N}_\mr{min.wgt}$ of the intermediate fusion network induced by these links.
We then find a maximum matching of $\mathcal{N}_\mr{min.wgt}$, which is the largest set of links that do not share any nodes, and contract these links in parallel.
By repeating this procedure until no links remain, we can determine the fusion order and calculate the resource overhead $Q$.
We illustrate an example in Fig.~\ref{fig:fusion_order}.
To compute a maximum matching, our software uses {\tt max\_weight\_matching} function from Python package \textit{NetworkX} \cite{aric2008networkx}, which is based on the algorithm in Ref.~\cite{galil1986efficient} and takes time of $\order{\abs{\text{number of nodes}}^3}$.

\subsubsection{Note on the average number of fusions}

One may want to use the average number of required fusion attempts to quantify resource overheads instead of the average number of required $\ket{\stargraph{3}}$ states.
In such a case, Eq.~\eqref{eq:contraction_weight_update} should be modified to
\begin{align*}
  w\qty(n_l) \coloneqq \begin{cases}
    \qty[w\qty(n_1) + w\qty(n_2) + 1]/{\psucc} & \text{if } n_1 \neq n_2, \\
    \qty[w\qty(n_1) + 1]/{\psucc} & \text{if } n_1 = n_2,
  \end{cases}
\end{align*}
and the weights of nodes should be initialized to~0.
All the other parts of the strategy remain the same.
\textit{\href{https://github.com/seokhyung-lee/OptGraphState}{OptGraphState}} provides an option to use this alternative resource measure instead of $Q$.

\subsection{Iteration \label{subsec:iteration}}

Since the method introduced above has randomness in several stages, it may produce a different outcome each time it is attempted.
We thus iterate the method a sufficient number of times and choose the best one.

Our software uses an \emph{adaptive} method to determine the iteration number:
Denoting the process to iterate the strategy $m$ times as $R(m)$, we first perform $R\qty(m_\mr{init})$ for a given integer $m_\mr{init} \geq 1$ and obtain $Q_\mr{opt}^{(1)}$, which is the minimal value of $Q$ obtained from the samples.
We then perform $R\qty(2m_\mr{init})$ and obtain $Q_\mr{opt}^{(2)}$.
If $Q_\mr{opt}^{(1)} \leq Q_\mr{opt}^{(2)}$, we stop the iteration and return $Q_\mr{opt}^{(1)}$.
If otherwise, we perform $R\qty(4m_\mr{init})$, obtain $Q_\mr{opt}^{(3)}$, and stop the iteration if $Q_\mr{opt}^{(2)} \leq Q_\mr{opt}^{(3)}$, which returns $Q_\mr{opt}^{(2)}$.
If $Q_\mr{opt}^{(2)} > Q_\mr{opt}^{(3)}$, we perform $R\qty(8m_\mr{init})$, obtain $Q_\mr{opt}^{(4)}$, and so on.
Here, we refer to this method as the \emph{adaptive iteration} with the given value of $m_\mr{init}$.

\revise{We emphasize that our strategy does not guarantee that the obtained generation method is strictly optimal.
As elaborated above, our approach involves simulating sufficiently many random samples and selecting the best performing one.}

\section{Applications of the strategy \label{sec:applications}}

In this section, we present the numerical results obtained by applying our strategy to various graphs.
We first analyze the distribution of resource overheads for random graphs, showing its tendency with respect to the numbers of vertices and edges.
We then provide numerical evidence indicating that each step of our strategy can significantly contribute to lowering the resource overhead.
\revise{Additionally, we} show the calculated resource overheads of various well-known graphs described in Sec.~\ref{subsec:graph_state}.
\revise{We lastly investigate the probability of successfully generating a graph state given a restricted number of available basic resource states and present the results for several graph states.}

Throughout the section, $\abs{V}$ and $\abs{E}$ for a given graph indicate the numbers of vertices and edges, respectively, and $\abs{E}_\mr{max}$ is defined as the maximal possible number of edges for a given value of $\abs{V}$ (under the assumption that there are no loops and multi-edges): $\abs{E}_\mr{max} = \abs{V}\qty(\abs{V} - 1)/2$.

\subsection{Analysis of random graphs \label{subsec:random_graph}}

\begin{figure}[!t]
  \centering
  \includegraphics[width=\columnwidth]{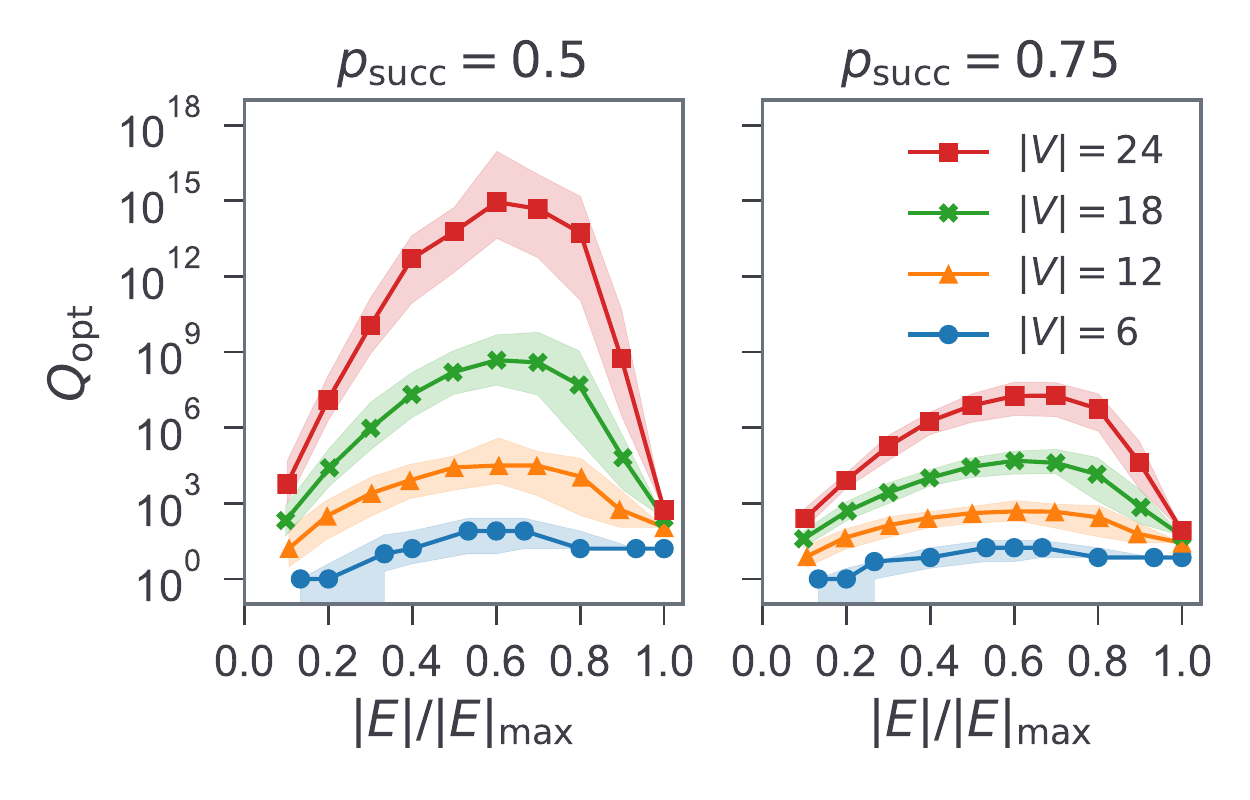}
  \caption{
    \textbf{Distribution of the optimized resource overhead $Q_\mathrm{opt}$ for random graphs.}
    Random graphs are sampled with fixed numbers of vertices ($\abs{V}$) and edges ($\abs{E}$) by the Erdős–Rényi model \cite{erdos1959random}.
    Two different fusion success rates are considered: $\psucc \in \qty{0.5, 0.75}$.
    $\abs{E}_\mr{max} = \abs{V}(\abs{V}-1)/2$ is the maximal possible number of edges for the given vertex number.
    For each combination of $\qty(p_\mathrm{succ}, \abs{V}, \abs{E})$, we sample 100 random graphs  and obtain the distribution of $Q_\mr{opt}$ through the adaptive iteration method of $m_\mr{init}=600$.
    The median of the distribution is indicated as a dot and its total range is shown as a shaded region.
  }
  \label{fig:random_graph_analysis}
\end{figure}

\begin{figure*}[!t]
  \centering
  \includegraphics[width=1.8\columnwidth]{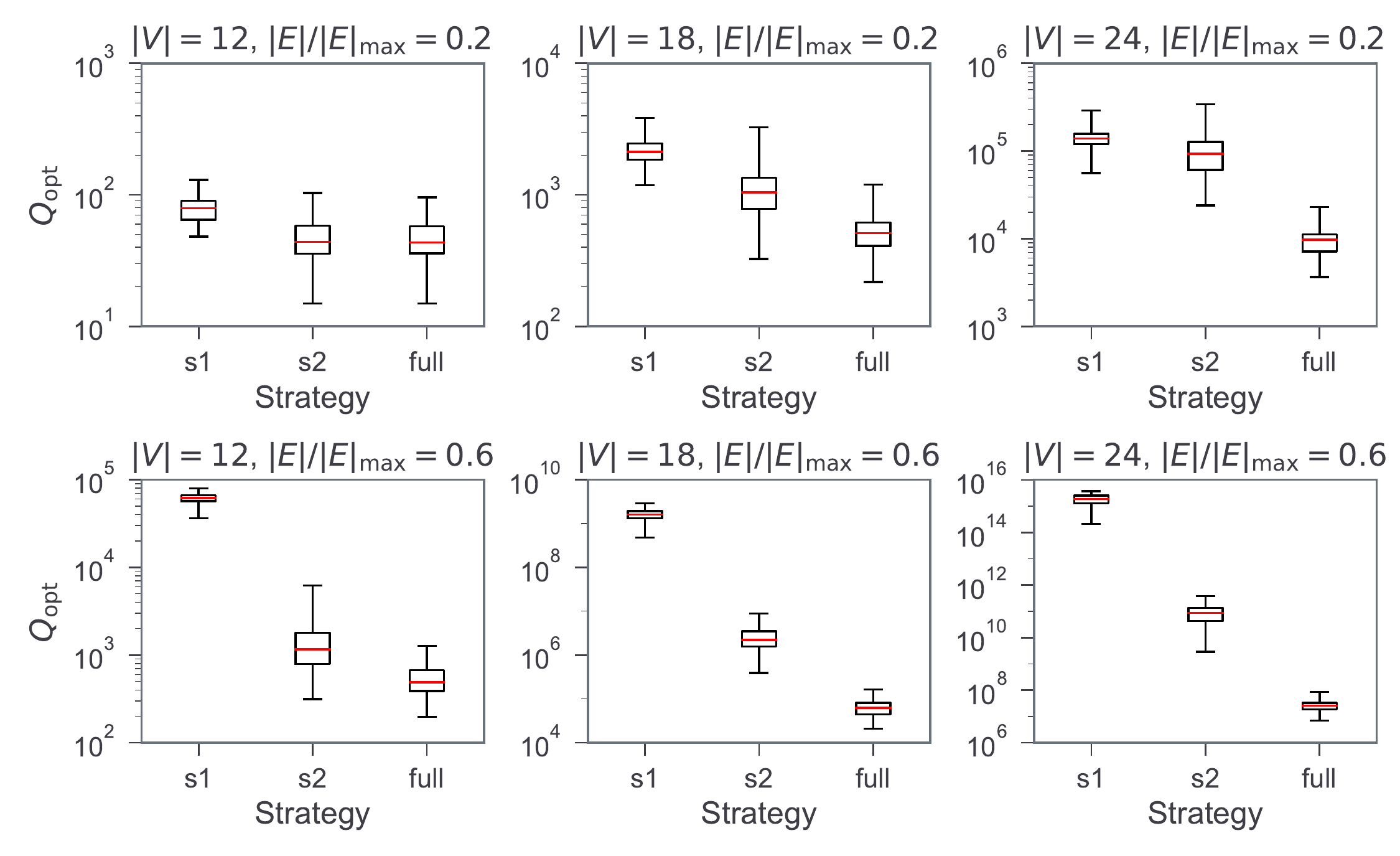}
  \caption{
    \textbf{Comparison of the distributions of the optimized resource overhead $Q_\mr{opt}$ for different optimization strategies.}
    Three strategies are considered: the strategy without unraveling (s1), the strategy with random selection of the fusion order (s2), and our full strategy.
    The subplots correspond to different values of $\abs{V} \in \qty{12, 18, 24}$ and $\abs{E}/\abs{E}_\mr{max} \in \qty{0.2, 0.6}$.
    For each setting, 100 graphs are sampled by the Erdős–Rényi model and 1200 iterations are done for each graph.
    (The adaptive iteration method is not used for fair comparisons.)
    The distribution of $Q_\mr{opt}$ is visualized as a box plot, where the red line indicates the median, the box extends from the first quartile (Q1) to the third quartile (Q3), and the whisker covers the entire range of the values.
  }
  \label{fig:performance_analysis}
\end{figure*}

To sample random graphs, we use the Erdős–Rényi model \cite{erdos1959random}, where all graphs that contain given fixed values of $\abs{V}$ and $\abs{E}$ have an equal probability.
Figure~\ref{fig:random_graph_analysis} visualizes the distributions of the obtained resource overheads optimized by our strategy for various values of $\abs{V}$ and $\abs{E}$ when $\psucc=0.5$ or $0.75$.
Here, we sample 100 random graphs for each combination $\qty(\psucc, \abs{V}, \abs{E})$ and use the adaptive iteration method of $m_\mr{init}=600$.
Several observations from the results are as follows:
\begin{itemize}
  \item $Q_\mr{opt}$ increases exponentially (or super-exponentially) as $\abs{V}$ grows when $\abs{E}/\abs{E}_\mathrm{max}$ is fixed.
  \item For a fixed value of $\abs{V}$, $Q_\mr{opt}$ is maximal when $\abs{E} \approx 0.6 \abs{E}_\mathrm{max}$. $Q_\mathrm{opt}$ is inversely correlated with $\abs{E}$ for large values of $\abs{E}$ since bipartitely-complete subgraphs and cliques are more likely to appear for when $\abs{E}$ is large.
  \item The fusion scheme with $\psucc=0.75$ may greatly reduce the order of $Q_\mr{opt}$, compared to the one with $\psucc=0.5$, especially when $\abs{V}$ is large. Note that, to achieve $\psucc=0.75$ with linear optics, we require an ancillary two-photon Bell state \cite{grice2011arbitrarily} or four ancillary unentangled photons \cite{ewert2014efficient} per fusion and photon-number resolving detectors that can discriminate at least four photons.
  On the other hand, the scheme with $\psucc=0.5$ requires only on-off detectors and no ancillary photons.
\end{itemize}

\subsection{Performance analysis \label{subsec:performance_analysis}}

We now show that our strategy is indeed effective by comparing it with two ``deficient'' strategies in which a certain stage is missing from the original ``full'' strategy.
In detail, we consider the following two alternative strategies:
\begin{enumerate}
  \item[(s1)] The strategy without the unraveling process, where the original graph is directly used for generating a fusion network. The other steps are the same as the full strategy.
  \item[(s2)] The strategy where the fusion order is randomly selected without using the min-weight-maximum-matching-first method. The other steps are the same as the full strategy.
\end{enumerate}
In Fig.~\ref{fig:performance_analysis}, the distributions of $Q_\mr{opt}$ optimized by these three strategies for random graphs are presented as box plots.
Each box extends from the first quartile (Q1) to the third quartile (Q3) and the corresponding whisker covers the entire range of the values.
It clearly shows that the full strategy is significantly more powerful than the deficient ones, especially when there exist many vertices and edges.
In other words, each step in the full strategy contributes to reducing the resource overhead.

\subsection{Applications to well-known graphs \label{subsec:well_known_graphs}}

\begin{figure}[!t]
  \centering
  \includegraphics[width=\columnwidth]{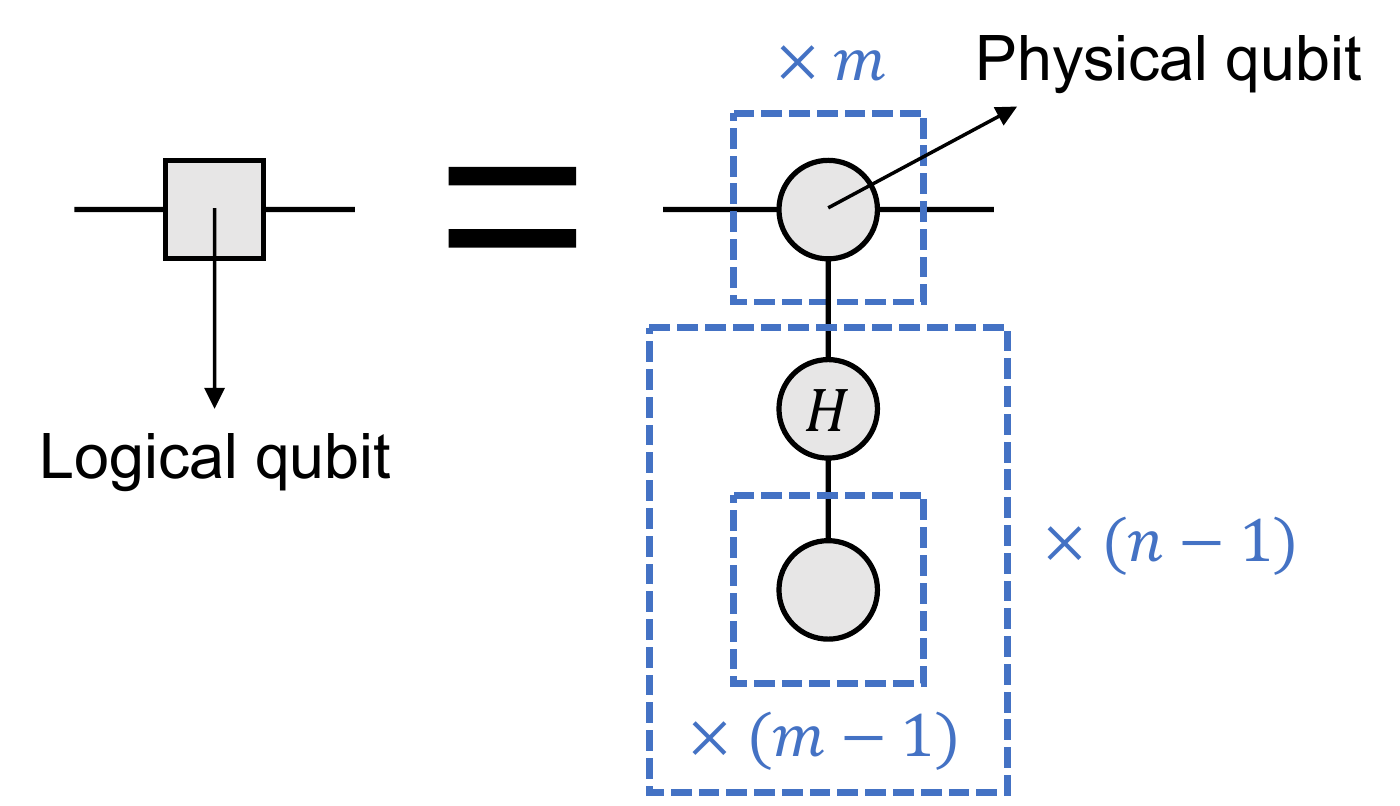}
  \caption{
    \textbf{Rule for converting an $(n, m)$ parity-encoded graph state into a physical-level graph state.}
    A dashed box with a text ``$\times N$'' (for an integer $N$) indicates a bundle of recurrent subgraphs.
    Namely, the subgraph inside the box is repeated $N$ times and, for each edge crossing the border of the box, an edge of the same format exists for every  repeated subgraph.
    See Ref.~\cite{lee2023parity} for more details.
  }
  \label{fig:encoded_graph}
\end{figure}

\begin{table*}[!tb]
\caption{
  \textbf{Results of resource overhead analyses for various well-known graphs.}
  See Fig.~\ref{fig:various_graphs} for the visualization of these graphs.
  $\abs{V}$ and $\abs{E}$ indicate the numbers of vertices and edges.
  $\abs{E}_\mr{max} = \abs{V}(\abs{V}-1)/2$ is the maximal possible number of edges.
  The optimized resource overheads $Q_\mr{opt}$ and the corresponding average numbers of fusion attempts are calculated for two fusion success rates: $\psucc \in \qty{0.5, 0.75}$.
  The adaptive iteration method of $m_\mr{init}=1200$ is used for the calculation.
}
\label{table:various_graphs_analysis}
\centering
\begin{tabular}{lrrrrrrr}
\toprule
\multirow{2}{*}{Graph} & \multirow{2}{*}{$\abs{V}$} & \multirow{2}{*}{$\abs{E}$} & \multirow{2}{*}{$\abs{E}/\abs{E}_\mr{max}$} & \multicolumn{2}{c}{$\psucc=0.5$} & \multicolumn{2}{c}{$\psucc=0.75$} \\ \cmidrule{5-8}
 &  &  &  & $Q_\mr{opt}$ & \#Fusions & $Q_\mr{opt}$ & \#Fusions \\ \midrule
6-vertex star ($G^{(6)}_*$) & 6 & 5 & 0.33 & $1.6 \times 10^1$ & $1.0 \times 10^1$ & $7.1 \times 10^0$ & $4.9 \times 10^0$ \\
12-vertex star ($G^{(12)}_*$) & 12 & 11 & 0.17 & $1.1 \times 10^2$ & $7.4 \times 10^1$ & $2.7 \times 10^1$ & $2.1 \times 10^1$ \\
18-vertex star ($G^{(18)}_*$) & 18 & 17 & 0.11 & $2.6 \times 10^2$ & $1.7 \times 10^2$ & $5.1 \times 10^1$ & $4.0 \times 10^1$ \\
24-vertex star ($G^{(24)}_*$) & 24 & 23 & 0.083 & $5.4 \times 10^{2}$ & $3.6 \times 10^{2}$ & $8.2 \times 10^{1}$ & $6.5 \times 10^{1}$ \\
$(3, 3)$-lattice & 9 & 12 & 0.33 & $5.4 \times 10^{2}$ & $3.7 \times 10^{2}$ & $5.5 \times 10^{1}$ & $4.6 \times 10^{1}$ \\
$(4, 4)$-lattice & 16 & 24 & 0.20 & $7.7 \times 10^{3}$ & $5.2 \times 10^{3}$ & $2.4 \times 10^{2}$ & $2.0 \times 10^{2}$ \\
$(5, 5)$-lattice & 25 & 40 & 0.13 & $1.0 \times 10^{5}$ & $6.7 \times 10^{4}$ & $9.9 \times 10^{2}$ & $8.2 \times 10^{2}$ \\
$(6, 6)$-lattice & 36 & 60 & 0.095 & $7.9 \times 10^{5}$ & $5.3 \times 10^{5}$ & $2.8 \times 10^{3}$ & $2.3 \times 10^{3}$ \\
$(1, 1, 1)$-RHG lattice & 18 & 24 & 0.16 & $1.9 \times 10^{4}$ & $1.3 \times 10^{4}$ & $3.9 \times 10^{2}$ & $3.4 \times 10^{2}$ \\
$(2, 2, 2)$-RHG lattice & 90 & 144 & 0.036 & $2.8 \times 10^{13}$ & $1.8 \times 10^{13}$ & $8.0 \times 10^{6}$ & $6.5 \times 10^{6}$ \\
$(2, 2)$-tree & 7 & 6 & 0.29 & $2.8 \times 10^{1}$ & $1.8 \times 10^{1}$ & $1.0 \times 10^{1}$ & $7.3 \times 10^{0}$ \\
$(2, 2, 2)$-tree & 15 & 14 & 0.13 & $2.1 \times 10^{2}$ & $1.4 \times 10^{2}$ & $4.0 \times 10^{1}$ & $3.1 \times 10^{1}$ \\
$(2, 2, 2, 2)$-tree & 31 & 30 & 0.065 & $1.6 \times 10^{3}$ & $1.1 \times 10^{3}$ & $1.4 \times 10^{2}$ & $1.1 \times 10^{2}$ \\
$(3, 3, 3)$-tree & 40 & 39 & 0.050 & $1.7 \times 10^{3}$ & $1.2 \times 10^{3}$ & $1.8 \times 10^{2}$ & $1.5 \times 10^{2}$ \\
$(4, 4, 4)$-tree & 85 & 84 & 0.024 & $1.2 \times 10^{4}$ & $7.8 \times 10^{3}$ & $6.1 \times 10^{2}$ & $4.9 \times 10^{2}$ \\
$(8, 2, 2)$-tree & 57 & 56 & 0.035 & $1.6 \times 10^{4}$ & $1.0 \times 10^{4}$ & $4.7 \times 10^{2}$ & $3.8 \times 10^{2}$ \\
Repeater graph with $m=3$ & 12 & 21 & 0.32 & $1.2 \times 10^{2}$ & $8.2 \times 10^{1}$ & $2.8 \times 10^{1}$ & $2.1 \times 10^{1}$ \\
Repeater graph with $m=4$ & 16 & 36 & 0.30 & $2.1 \times 10^{2}$ & $1.4 \times 10^{2}$ & $4.3 \times 10^{1}$ & $3.3 \times 10^{1}$ \\
Repeater graph with $m=6$ & 24 & 78 & 0.28 & $5.4 \times 10^{2}$ & $3.6 \times 10^{2}$ & $8.2 \times 10^{1}$ & $6.5 \times 10^{1}$ \\
$(2, 2)$ parity-encoded 3-star & 12 & 17 & 0.26 & $1.2 \times 10^{2}$ & $8.2 \times 10^{1}$ & $2.8 \times 10^{1}$ & $2.1 \times 10^{1}$ \\
$(3, 3)$ parity-encoded 3-star & 27 & 48 & 0.14 & $8.8 \times 10^{2}$ & $5.9 \times 10^{2}$ & $1.1 \times 10^{2}$ & $8.4 \times 10^{1}$ \\
$(4, 4)$ parity-encoded 3-star & 48 & 95 & 0.084 & $2.4 \times 10^{3}$ & $1.6 \times 10^{3}$ & $2.3 \times 10^{2}$ & $1.9 \times 10^{2}$ \\
$(5, 5)$ parity-encoded 3-star & 75 & 158 & 0.057 & $1.0 \times 10^{4}$ & $6.8 \times 10^{3}$ & $5.3 \times 10^{2}$ & $4.3 \times 10^{2}$ \\
$(2, 2)$ parity-encoded 6-cycle & 24 & 42 & 0.15 & $1.3 \times 10^{3}$ & $8.5 \times 10^{2}$ & $1.2 \times 10^{2}$ & $9.9 \times 10^{1}$ \\
$(3, 3)$ parity-encoded 6-cycle & 54 & 114 & 0.080 & $6.7 \times 10^{3}$ & $4.4 \times 10^{3}$ & $3.9 \times 10^{2}$ & $3.1 \times 10^{2}$ \\
$(4, 4)$ parity-encoded 6-cycle & 96 & 222 & 0.049 & $2.1 \times 10^{4}$ & $1.4 \times 10^{4}$ & $8.8 \times 10^{2}$ & $7.1 \times 10^{2}$ \\

\bottomrule
\end{tabular}
\end{table*}

We here investigate the resource overheads of the graph states in Sec.~\ref{subsec:graph_state}, which are utilized in various quantum tasks such as MBQC, FBQC, quantum repeaters, and quantum error correction.
Besides them, we also consider parity-encoded graph states, which are used for basic resource states of parity-encoding-based topological quantum computing (PTQC) protocol in Ref.~\cite{lee2023parity} and FBQC in Ref.~\cite{bartolucci2023fusion}.
The \emph{$(n, m)$ parity code} (or \emph{generalized Shor code}) \cite{ralph2005loss} encodes a single logical qubit with the basis of
\begin{align*}
  \qty{\qty(\ket{0}^{\otimes m} + \ket{1}^{\otimes m})^{\otimes n} \pm \qty(\ket{0}^{\otimes m} - \ket{1}^{\otimes m})^{\otimes n}},
\end{align*}
where $\qty{\ket{0}, \ket{1}}$ is the physical-level basis.
An \emph{$(n, m)$ parity-encoded graph state} indicates a graph state in which the qubits on the vertices are encoded with the $(n, m)$ parity code.
Such an encoded graph state can be rewritten as a graph state of physical-level qubits according to the rule in Fig.~\ref{fig:encoded_graph} \cite{lee2023parity}.
We cover two types of logical-level graphs in the calculation: the 3-vertex star graph (for PTQC \cite{lee2023parity}) and 6-vertex cycle graph (for FBQC \cite{bartolucci2023fusion}).

In Table~\ref{table:various_graphs_analysis}, we list the results of the resource analyses for these graph states, together with the basic information of the graphs.
Additionally, in Appendix~\ref{app:examples}, we present several \revise{explicit} examples of the application of our strategy with visualization.
\revise{
We note that the extremely high resource overheads of RHG lattices do not necessarily render them impractical.
From the perspective of fault-tolerant quantum computing \cite{raussendorf2006fault}, certain levels of absent vertices or edges resulting from fusion failures can be endured \cite{barrett2010fault,auger2018fault}.
Hence, many previous schemes \cite{li2015resource,omkar2022all,lee2023parity,bartolucci2023fusion} take an approach of initially generating ``unit'' resource states successfully, followed by merging them while allowing fusion failures.
In this context, our strategy can be utilized to evaluate the resource costs of these unit states.
For example, they can be tree graphs \cite{li2015resource}, parity-encoded 3-star graphs \cite{omkar2022all,lee2023parity}, or parity-encoded 6-cycle graphs \cite{bartolucci2023fusion}, all of which are presented in Table~\ref{table:various_graphs_analysis}.
}

\subsection{\revise{Success probability of graph state generation} \label{subsec:generation_success_prob}}
\revise{We emphasize that the resource overhead $Q$ we have used so far is defined by the \emph{expected value} of the resource count required to generate the desired graph state $\ket{G}$.
Namely, if we define a discrete random variable $C$ by this resource count, then
\begin{align*}
  Q = \Exp{C} = \sum_{c=1}^\infty c\Pr(C=c).
\end{align*}
One may want to know more information on the probability mass function (PMF) $\Pr(C=c)$, not just its expectation value.
Moreover, the corresponding cumulative mass function (CMF) indicates the probability $P_\mr{succ}(c)$ of the successful generation of $\ket{G}$ when $c$ resource states are provided, namely,
\begin{align}
  P_\mr{succ}(c) = \sum_{c'=1}^c \Pr\qty(C=c'), \label{eq:cmf_def}
\end{align}
which may be a more practical indicator than $Q$ to assess the performance of a scheme.

We first investigate the case where two graph states with resource counts $C_1$ and $C_2$ are merged by a single fusion of success rate $p_\mr{s}$ to form a graph state with resource count $C_3$.
For each $i \in \qty{1,2,3}$, let us define $q_i(\cdot)$ by the probability distribution function (PDF) corresponding to the PMF $\Pr\qty(C_i = c)$; namely,
\begin{align*}
  q_i(x) \coloneqq \sum_{c=1}^\infty \Pr\qty(C_i = c)\delta(x-c),
\end{align*}
where $\delta$ is the Kronecker delta function.
Then $q_1$, $q_2$, and $q_3$ are related as
\begin{align}
  &q_3 = p_\mr{s} \qty(q_1 \ast q_2) + p_\mr{s}(1-p_\mr{s})\qty(q_1 \ast q_1 \ast q_2 \ast q_2) \nonumber \\ 
  &+ p_\mr{s}(1-p_\mr{s})^2 \qty(q_1 \ast q_1 \ast q_1 \ast q_2 \ast q_2 \ast q_2) \nonumber\\ 
  &+ p_\mr{s}(1-p_\mr{s})^3 \qty(q_1 \ast q_1 \ast q_1 \ast q_1 \ast q_2 \ast q_2 \ast q_2 \ast q_2) \nonumber\\ 
  &+ \cdots, \label{eq:pdf_relation}
\end{align}
where the ``$\ast$'' symbol indicates convolution defined as
\begin{align*}
  (f \ast g)(x) \coloneqq \int_{-\infty}^{\infty} f(t) g(x-t) \dd t.
\end{align*}
Here, each term including $p_\mr{s}(1-p_\mr{s})^l$ for $l \geq 0$ corresponds to the case where the fusion fails $l$ times and succeeds on the next attempt.
The convolution theorem \cite{arfken2011mathematical} states that the Fourier transformation (FT) defined as
\begin{align*}
  \FT{f}(k) = \tilde{f}(k) \coloneqq \int_{-\infty}^\infty f(x) e^{ikx} \dd x
\end{align*}
converts convolution into multiplication of functions as
\begin{align*}
  \FT{f \ast g}(k) = \tilde{f}(k) \tilde{g}(k).
\end{align*}
Thus, applying the FT to the both sides of Eq.~\eqref{eq:pdf_relation}, we obtain
\begin{align*}
  \tilde{q}_3(k) &= \sum_{l=0}^\infty p_\mr{s}(1-p_\mr{s})^l \qty[\tilde{q}_1(k) \tilde{q}_2(k)]^{l+1} \\ 
  &= \frac{p_\mr{s} \tilde{q}_1(k) \tilde{q}_2(k)}{1-(1-p_\mr{s})\tilde{q}_1(k) \tilde{q}_2(k)},
\end{align*}
which gives
\begin{align}
  \tilde{q}_3(k)^{-1} &= \frac{1}{p_\mr{s}} \tilde{q}_1(k)^{-1} \tilde{q}_2(k)^{-1} - \frac{1-p_\mr{s}}{p_\mr{s}}. \label{eq:ftpdf_update_rule}
\end{align}
Therefore, considering that the PDF of the resource count for the basic resource state $\ket{\stargraph{3}}$ is $\delta(x-1) \eqqcolon q_\mr{base}(x)$, the Fourier-transformed PDF (FTPDF) for every graph state can be written as the inverse of a polynomial of
\begin{align*}
  \tilde{q}_\mr{base}^{-1}(k) = e^{-ik} \eqqcolon z^{-1}.
\end{align*}
Namely, for every FTPDF $\tilde{q}(k)$, there exists a series of real numbers $\qty{a_l}_{l=0}^L$ that satisfies
\begin{align}
  \tilde{q}(k) = \frac{1}{\sum_{l=0}^L a_l z^{-l}} = \frac{1}{\sum_{l=0}^L a_l e^{-ikl}}.
  \label{eq:ftpdf_poly_expr}
\end{align}
To compute the FTPDF for a desired graph state, we just need to assign the PDF of $z$ to every node of the fusion network and sequentially apply the rule of Eq.~\eqref{eq:ftpdf_update_rule} for every link contraction.

After obtaining the final FTPDF $\tilde{q}(k)$ in the form of Eq.~\eqref{eq:ftpdf_poly_expr}, we need to recover the PMF $\Pr(C=c)$ from it.
By applying the Taylor expansion at $z=0$ on
\begin{align*}
  \tilde{q}(k) &= \frac{z^{L}}{\sum_{l=0}^{L} a_l z^{L-l}} = \frac{z^L}{\sum_{l=0}^{L} a_{L-l} z^l},
\end{align*}
we get the power series
\begin{align*}
  \tilde{q}(k) = z^L \sum_{j=0}^\infty b_j z^j,
\end{align*}
where $\qty{b_j}$ is recurrently defined as
\begin{align*}
  b_0 &= a_L^{-1}, \\ 
  b_j &= -\sum_{j'=\max(0,j-L)}^{j-1} \frac{a_{L-j+j'}}{a_L} b_{j'}.
\end{align*}
Since $z^j = e^{ikj}$ is the FT of $\delta(x - j)$, we conclude that the PMF is
\begin{align*}
  \Pr(C=c) = \begin{cases}
    0 & \text{if } c < L, \\
    b_{c-L} & \text{otherwise}.
  \end{cases}
\end{align*}
Additionally, we obtain the CMF $P_\mr{succ}(c)$ in Eq.~\eqref{eq:cmf_def} as
\begin{align*}
  P_\mr{succ}(c) = \begin{cases}
    0 & \text{if } c < L, \\
    d_{c-L} & \text{otherwise},
  \end{cases}
\end{align*}
where
\begin{align*}
  d_0 &= a_L^{-1}, \\ 
  d_j &= \sum_{j' = \max(0, j-L-1)}^{j-1} \frac{a_{L-j+j'+1} - a_{L-j+j'}}{a_L} d_{j'},
\end{align*}
if we define $a_{-1} \coloneqq 0$.

\begin{figure}[!t]
  \centering
  \includegraphics[width=\linewidth]{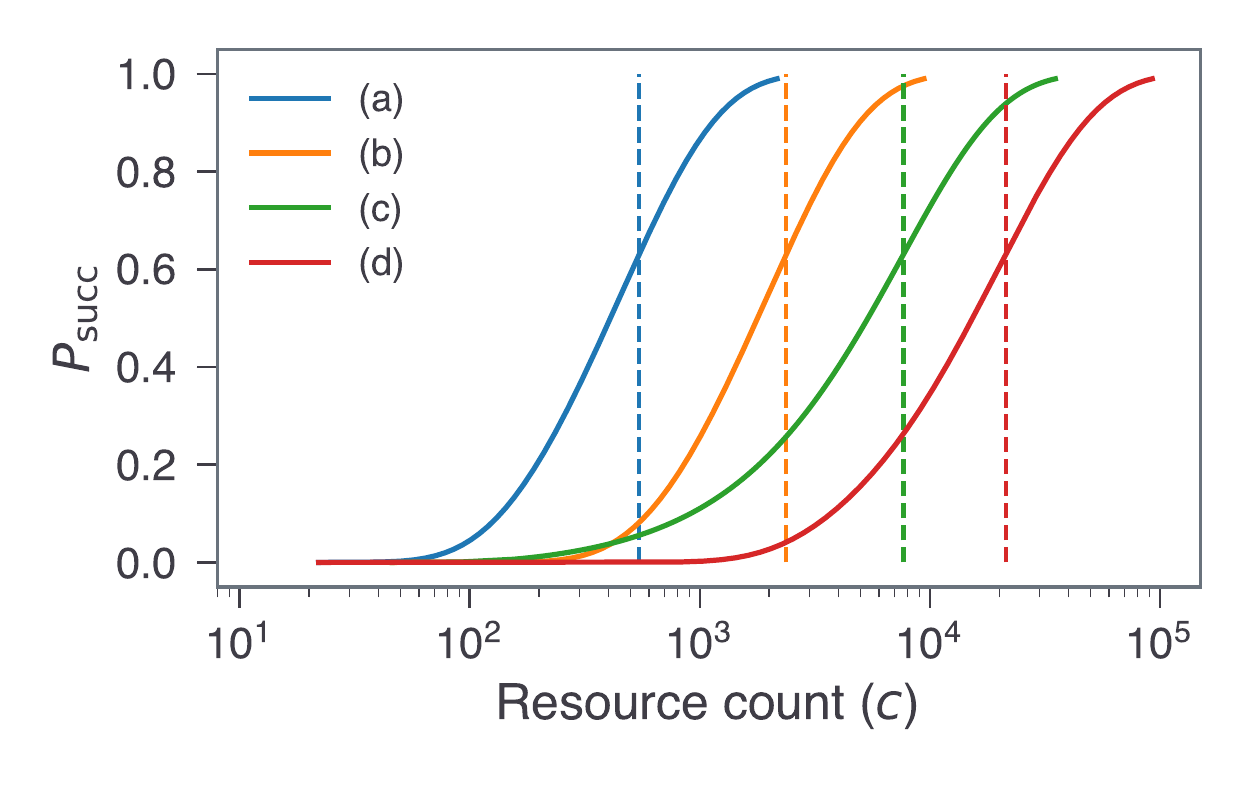}
  \caption{
    \revise{\textbf{Success probabilities of graph state generation as functions of the number of provided basic resource states.} 
    The solid lines respectively indicate the success probabilities $P_\mr{succ}(c)$ for the graph states of \textbf{(a)} the repeater graph with $m=6$, \textbf{(b)} $(4, 4)$ parity-encoded 3-star, \textbf{(c)} $(4,4)$-lattice, and \textbf{(d)} $(4,4)$ parity-encoded 6-cycle. 
    Each dashed line represents the corresponding resource overhead $Q$, which is the expected value of the resource count when $P_\mr{succ}$ is the cumulative mass function.}
  }
  \label{fig:generation_success_rate}
\end{figure}

The above calculations are implemented in our software \emph{OptGraphState}.
In Fig.~\ref{fig:generation_success_rate}, we display the computed success probabilities $P_\mr{succ}$ for generating various graph states, plotted against the resource count $c$.
Lines labeled (a)--(d) correspond to the repeater graph with $m=6$, $(4,4)$ parity-encoded 3-star graph, $(4,4)$-lattice graph, and $(4,4)$ parity-encoded 6-cycle graph, respectively.
We note that the resource overhead $Q$, which is shown as dashed lines, matches the resource count when the associated success probability is roughly 60\%.
For achieving a greater success probability, such as 90\%, one would need approximately twice as many resource states as $Q$.}

\section{Remarks \label{sec:remarks}}

Graph states are versatile resource states for various tasks on quantum computation and communication, such as \revise{MBQC} \cite{raussendorf2001one,raussendorf2006fault}, \revise{FBQC} \cite{bartolucci2023fusion}, quantum error correction \cite{schlingemann2001quantum,pirker2017construction}, and quantum repeaters \cite{zwerger2012measurement}.
However, in optical systems, the non-deterministic nature of entangling operations hinders the generation of large-scale graph states; thus, the generation process should be carefully designed.

In this work, we \revise{introduced} a graph-theoretical strategy to construct a resource-efficient method for generating an arbitrary graph state with the type-II fusion operation.
Here, the resource overhead is quantified by the average number of required basic resource states (three-qubit star graph states) to generate the graph state without failed fusions.
As outlined in Sec.~\ref{sec:strategy}, the strategy is composed of multiple trials to find the optimal one, where each round contains three stages: unraveling the graph, constructing a fusion network, and determining the fusion order.
In Sec.~\ref{sec:applications}, we applied the strategy to various graph states and verified numerically that each step of the strategy is indeed necessary to achieve high resource efficiency.
\revise{Moreover, we described a recursive technique to determine the success probability of generating a graph state as a function of the resource cost and tested it on several representative graph states.}

We anticipate that our strategy and software will aid researchers in designing experimentally feasible approaches utilizing photonic graph states and in evaluating the practicality of their proposed schemes.
For example, the basic resource states of MBQC and FBQC can be logically-encoded star or cycle graph states \cite{lee2023parity,bartolucci2023fusion}.
Employing larger or more complex codes may improve the fault-tolerance of these schemes; however, generating such resource states could become a bottleneck in their implementation.
Our strategy can contribute to evaluating such a trade-off relation and identifying the most practical sweet spot.

We lastly note several interesting unsolved problems related to our work:
\begin{enumerate}
  \item \textbf{Generalization of unraveling.} For a given graph state $\ket{G}$, how can we identify another graph state $\ket{G'}$ such that $\ket{G}$ can be generated from $\ket{G'}$ using a combination of fusions, single-qubit Clifford (or general) operations, single-qubit measurements, and classical communications, resulting in a reduction of the overall resource overhead? This problem bears similarities to the equivalence problem of graph states \cite{nest2004graphical,nest2004efficient,dahlberg2018transforming}, but fusions are included as allowable operations and resource overheads for fusion-based generation are considered.
  \item \textbf{Lower bound of resource overhead.} Is it possible to find a (sufficiently tight) lower bound of the resource overhead $Q$? If such a lower bound can be computed, it would enable us to assess whether the resource overhead optimized by our strategy is indeed close to the real optimal value.
  \item \textbf{Behavior of $Q_\mr{opt}$ against $\abs{E}/\abs{E}_\mr{max}$.} In Fig.~\ref{fig:random_graph_analysis}, $Q_\mr{opt}$ exhibits an intriguing behavior, where it is maximized around $\abs{E}/\abs{E}_\mr{max}=0.6$ regardless of $\abs{V}$. Can it be explained analytically? Is $Q_\mr{opt}$ related to a specific property of the graph or graph state, such as the multipartite entanglement of the graph state \cite{hein2004multiparty}?
  \item \revise{\textbf{Usage of larger basic resource states.} Using a graph state larger than the three-qubit star graph state $\ket{\stargraph{3}}$ as the basic resource state can be beneficial. While preparing larger basic resource state might be challenging, it can reduce the resource overhead. We anticipate that the reduction would be approximately proportional to the original overhead of the new basic resource state. For instance, employing $\ket{\stargraph{4}}$ could lead to a fourfold reduction in overhead. If $\ket{\stargraph{n}}$ is used as the basic resource state, only minor adjustments are needed in our strategy: Within each node group of a fusion network shown in Fig.~\ref{fig:fusion_network}, contract adjacent nodes beforehand so that each node represents a star graph state with at most $n$ qubits. However, if the basic resource state is not a star graph state, it might necessitate a complete overhaul of the algorithm, which will be worth investigating.
  }
  \item \revise{\textbf{Tolerance for fusion failures.} The aim of our strategy is to generate a graph state without any failed fusions. However, in practical scenarios, we may consider allowing some fusion failures at the cost of some missing vertices and edges from the lattice, which can be tolerable depending on the characteristics and usage of the generated state. For example, in parity-encoded graph states, which are graph states of logically encoded qubits, such defects may lead to correctable errors. If we allow a degree of failed fusions, how many vertices or edges would be missing from the resulting lattice? In such cases, how can we determine an efficient generation scheme and calculate its resource overhead? Can we relate the fault-tolerance of a graph state and the resource overhead to generate it?}
\end{enumerate}

\section*{Acknowledgement}

This work was supported by the National Research Foundation of Korea (NRF) grants funded by the Korean government (Grant Nos. NRF-2020R1A2C1008609, NRF-2023R1A2C1006115, and NRF-2022M3K4A1097117) via the Institute of Applied Physics at Seoul National University, by the Institute of Information \& Communications Technology Planning \& Evaluation (IITP) grant funded by the Korea government (MSIT) (IITP-2021-0-01059 and IITP-2023-2020-0-01606).

\bibliographystyle{quantum}
\bibliography{references}

\onecolumn\newpage
\appendix

\section{Examples of the application of the strategy \label{app:examples}}

\begin{figure*}[!ht]
  \centering
  \includegraphics[width=\textwidth]{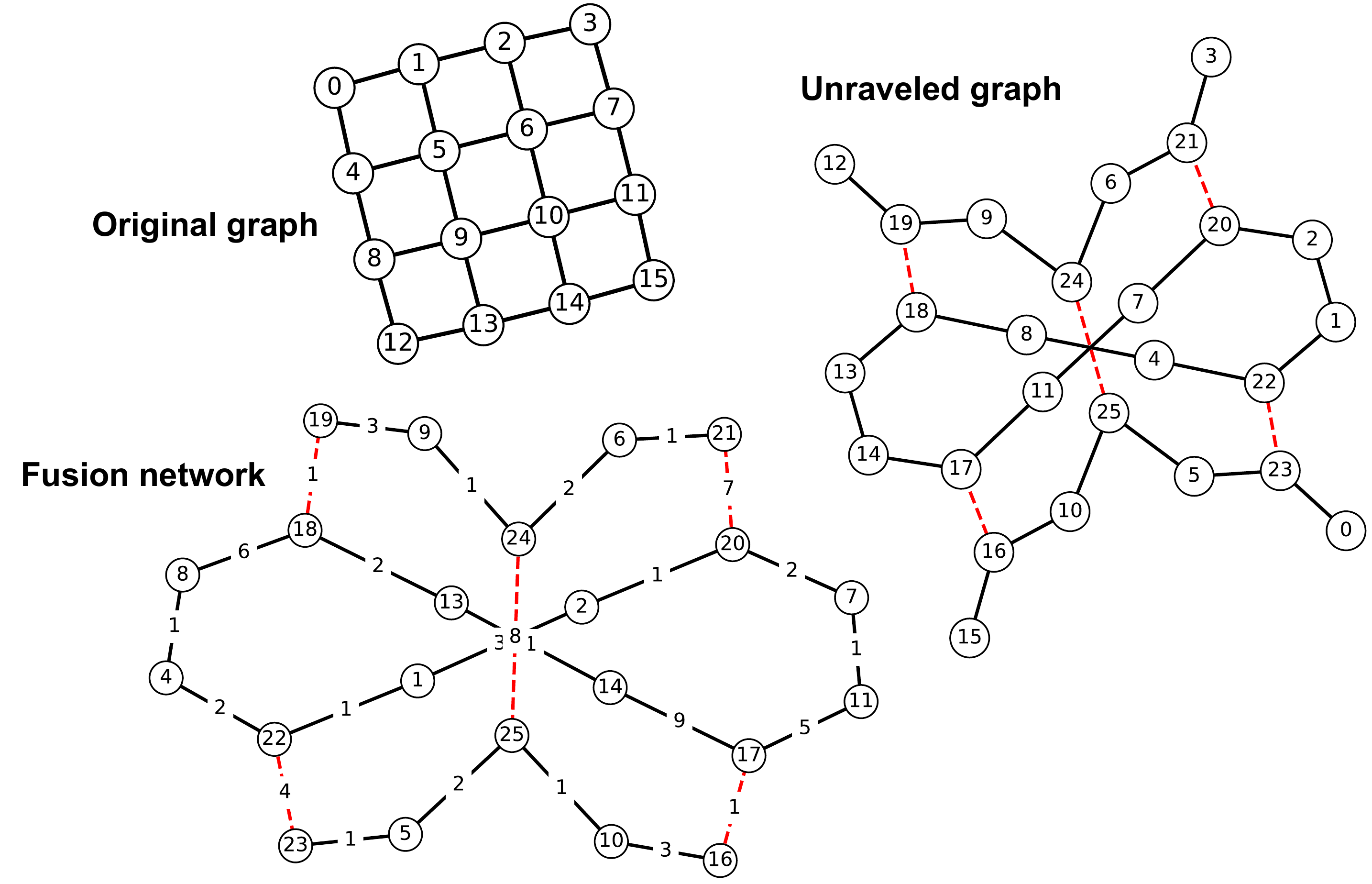}
  \caption{\textbf{$(4, 4)$-lattice graph and its unraveled graph and fusion network obtained by our strategy.} The unraveled graph and fusion network give the resource overhead of $7680$ when $\psucc=0.5$.
    See Appendix~\ref{app:examples} for their interpretation.}
  \label{fig:example_lattice}
\end{figure*}

\begin{figure*}[!ht]
  \centering
  \includegraphics[width=\textwidth]{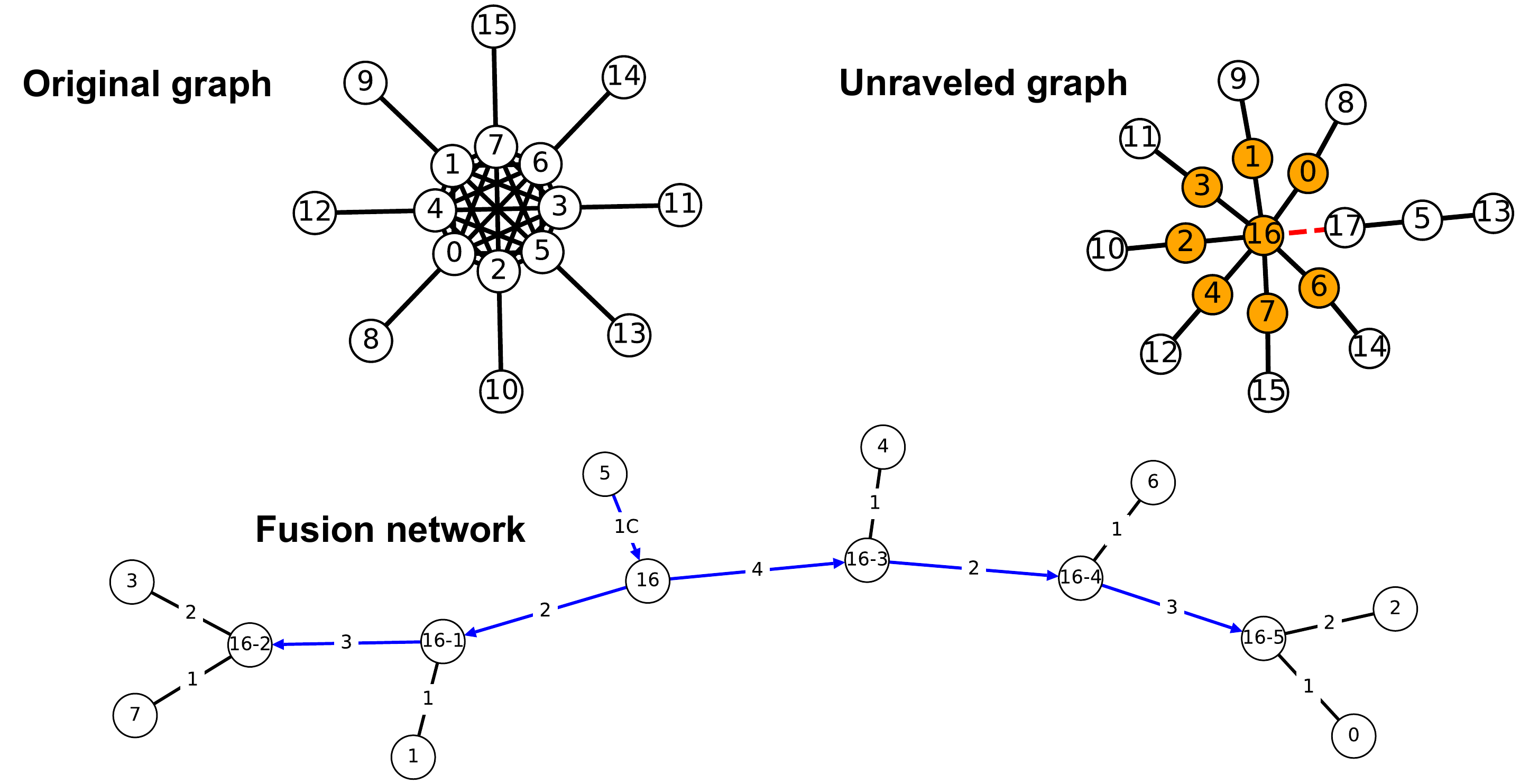}
  \caption{
        \textbf{Repeater graph with $m=4$ and its unraveled graph and fusion network obtained by our strategy.} 
        The unraveled graph and fusion network give the resource overhead of $208$ when $\psucc=0.5$.
        See Appendix~\ref{app:examples} for their interpretation.
        In the unraveled graph, the qubits at the orange vertices are subjected to non-trivial Clifford gates: $R_X(\pi/2) \coloneqq \exp\qty[i(\pi/4)\hat{X}]$ to qubit 16 and $R_Z(\pi/2) \coloneqq \exp\qty[i(\pi/4)\hat{Z}]$ to the others.
        In the fusion network, the fusion of the link connecting the nodes 5 and 16 is accompanied by $R_X(\pi/2)$ applied to the qubit in node 16 before the fusion is performed.
    }
  \label{fig:example_repeater}
\end{figure*}

\begin{figure*}[!ht]
  \centering
  \includegraphics[width=\textwidth]{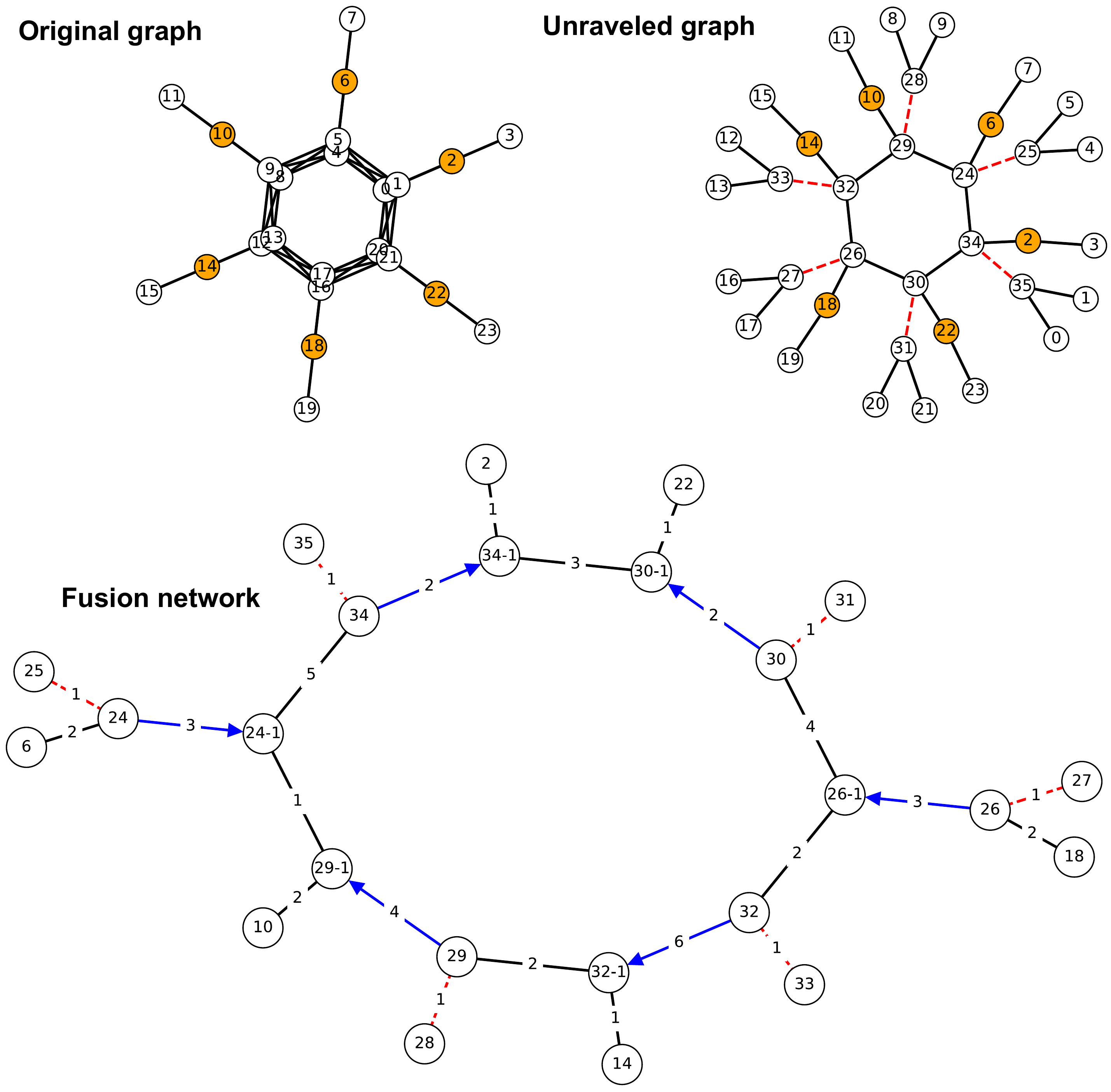}
  \caption{
        \textbf{$(2, 2)$ parity-encoded 6-cycle graph and its unraveled graph and fusion network obtained by our strategy.} 
        The unraveled graph and fusion network give the resource overhead of $1280$ when $\psucc=0.5$.
        See Appendix~\ref{app:examples} for their interpretation.
        In the original and unraveled graphs, the qubits at the orange vertices are subjected to the Hadamard gate.
    }
  \label{fig:example_parity}
\end{figure*}

In this Appendix, we present examples of applying our strategy to several graphs, which is obtained by using our Python package \textit{\href{https://github.com/seokhyung-lee/OptGraphState}{OptGraphState}}.
Figures~\ref{fig:example_lattice}, \ref{fig:example_repeater}, and \ref{fig:example_parity} show the \textit{$(4, 4)$-lattice graph}, \textit{repeater graph with $m=4$}, and \textit{$(2, 2)$ parity-encoded 6-cycle graph} and their unraveled graphs and fusion networks that give the resource overheads in Table~\ref{table:various_graphs_analysis} when $\psucc=0.5$, which are plotted by using the \textit{python-igraph} library.
The description of the various elements of these figures is as follows.

\paragraph*{Original graph:}
\begin{itemize}
    \item A number inside each vertex is its unique name.
    \item Orange vertices indicate qubits with non-trivial Clifford gates.
\end{itemize}

\paragraph*{Unraveled graph:}
\begin{itemize}
    \item A number inside each vertex is its unique name. If a vertex is originated from a vertex in the original graph, they have the same name.
    \item Black solid lines are edges of the unraveled graph and red dashed lines indicate external fusions.
    \item Orange vertices indicate qubits with non-trivial Clifford gates.
\end{itemize}

\paragraph*{Fusion network:}
\begin{itemize}
    \item A number inside each node is its unique name. Each seed node has the same name as the vertex in the unraveled graph that the node is originated from; namely, the qubit at the vertex is the root qubit of the $\ket{\stargraph{3}}$ state of the seed node. Non-seed nodes have names like `$i$-$j$', where $i$ is the name of the seed node in the same node group and $j$ is an index starting from $1$.
    \item Black solid lines are leaf-to-leaf links, red dashed lines are root-to-root links, and blue arrows are root-to-leaf links. Each arrow for a root-to-leaf link points from the node that contains the leaf qubit involved in the fusion to the other node.
    \item A number placed on each link indicates the order of the fusion. The fusions should be performed in the order in which these numbers increase and those with the same number can be done simultaneously.
    \item If the text placed on a link ends with the letter `C' (such as `1C'), it means that the corresponding fusion is accompanied by non-trivial Clifford gates applied to one or both of the qubits before the fusion is performed.
\end{itemize}

\end{document}